\documentclass[preprint,prd,aps,showpacs,showkeys,nofootinbib]{revtex4}
\usepackage{graphicx}
\usepackage{amsmath}
\usepackage{hyperref}
\usepackage[T1]{fontenc}
\usepackage{soul}
\usepackage{color,xcolor}
\begin{document}

	
	\title{Transition magnetic moment of Majorana neutrinos in the triplets
		next-to-minimal MSSM}
	
	\author{Zhao-Yang Zhang$^1$, Jin-Lei Yang$^{2,3}$\footnote{jlyang@hbu.edu.cn}, Hai-Bin Zhang$^{2,3}$\footnote{hbzhang@hbu.edu.cn},
		Tai-Fu Feng$^{1,2,3,4}$\footnote{fengtf@hbu.edu.cn}}
	
	\affiliation{$^1$Department of Physics, Guangxi University, Nanning, 530004, China\\
		$^2$Department of Physics, Hebei University, Baoding, 071002, China\\
		$^3$Key Laboratory of High-precision Computation and Application of Quantum Field Theory of Hebei Province, Baoding, 071002, China\\
		$^4$Department of Physics, Chongqing University, Chongqing 401331, China}

	\begin{abstract}
		The TNMSSM is an attractive extension of the Standard Model. It combines the advantages of the NMSSM and the TMSSM to give three tiny Majorana neutrinos masses via a type I+II seesaw mechanism. With the on-shell renormalization scheme, we consider the neutrino masses up to one loop approximation. Applying the effective Lagrangian method, we study the transition magnetic moments of Majorana neutrinos and consider the normal hierarchy (NH) and inverse hierarchy (IH) neutrino mass spectra within the constraints of experimental data on neutrino oscillations. The solar neutrino transition magnetic moment is further deduced, and compared with the XENONnT experiment limit.

	\end{abstract}

	\maketitle
	\section{Introduction\label{sec1}}
	\indent\indent
	From recent neutrino oscillation experiments, it is known that neutrinos have nonzero masses and mix with each other (see refs.~\cite{neu-b1,neu-b2,neu-b3,neu-b4,neu-b5}). However, the mass and mixing cannot be accounted for explicitly in the Standard Model (SM), so it is needed to extend the SM to fit the neutrino oscillation experiments.
	 The Minimal Supersymmetric Extension of the SM (MSSM) is a relatively simple extension of the SM, but it does not account for neutrino masses and does not provide a perfect explanation of the $\mu$ problem (\cite{up p1}) and the hierarchy problem(\cite{hire p1, hire p2}). In the light of this reality, the next-to-minimal supersymmetric of the Standard Model (NMSSM) emerged \cite{NMS p1,NMS p2}. It introduces a singlet with the hypercharge Y equal to 0, solving the $\mu$ problem. However,
	 NMSSM is no better at improving little hierarchy problem. Because the interaction of singlet states produces additional Higgs quartic, this can solve the small hierarchy problem in the NMSSM. But the additional Higgs quartic directly contributes to the Higgs mass,  while the additional Higgs quartic is suppressed in the large $\tan\beta$ limit\cite{UH p1,UH p2,UH p3,UH p4}. Extending the MSSM by adding the $SU(2)$ triplets (TMSSM)  \cite{TM p1,TM p2,TM p3} can provide such a Higgs quartic naturally. This model becomes more attractive when the triplet state has a nonzero hypercharge. In this case, the quartic couplings for the Higgs boson are not suppressed in the large $\tan\beta$ limit. Based on the above advantages, the next-to-minimal Supersymmetric Standard Model with triplets (TNMSSM) \cite{TNM p1} is an attractive extension of the SM.

	The TNMSSM includes one singlet with zero hypercharge, two triplet states with hypercharges  ${\pm1}$, and introduce three right-handed neutrinos with zero hypercharge. The right-handed neutrinos couple to the singlet state, and the left-handed neutrinos couples to the triplet state T. When the singlet scalar (Higgs) and triplet states acquire VEV, the right-handed neutrinos and the left-handed neutrinos acquire Majorana masses. Combining the Majorana mass terms with the Dirac mass terms, the tiny neutrino masses can be obtained by the type I+II seesaw mechanism \cite{type1 p1,type2 p1}.
	
	In general the transformation between the mass eigenstates of neutrinos $\nu_{1,2,3}$ and the flavor eigenstates $\nu_{e,\mu,\tau}$ is described via the Pontecorvo–Maki–Nakagawa–Sakata matrix $U_{\rm {PMNS}}$ \cite{UPMNS p1,UPMNS p2}. The constituent parameters of the $U_{\rm {PMNS}}$ matrix and the squared mass differences have been well measured by the neutrino oscillation experiments, the results read \cite{MASS DATA p1}:
	$$ \Delta m_{\nu_{21}}^2=(7.19-7.60) \times 10^{-5} \;{\rm eV}^2, $$
	$$ | \Delta m_{\nu_{32}}^2 |_\text{NH}=(2.42-2.48)\times 10^{-3} \;{\rm eV}^2 , $$
	$$ | \Delta m_{\nu_{32}}^2 |_\text{IH}=(2.48-2.54)\times 10^{-3} \;{\rm eV}^2 , $$
	$$ \text{sin}^2\theta_{12}=0.30-0.32 , $$
	$$ \text{sin}^2\theta_{23}=0.525-0.578 , $$ 
	\begin{equation}
		\text{sin}^2\theta_{13}=0.0217-0.0230\label{data}.
	\end{equation}
	where $| \Delta m_{\nu_{32}}^2 |_\text{NH}$ is for normal hierarchy (NH), $ | \Delta m_{\nu_{32}}^2 |_\text{IH}$ is for inverse hierarchy (IH). 	Eq.
	(1) shows that the three massive neutrinos are not degenerate, which indicates that nonzero
	neutrino transitions can take place by the electroweak radiation effects. The transition magnetic
	moments are one of the most important characteristics of massive Majorana neutrinos. Since neutrinos are detected indirectly, one of the more effective ways to study neutrino properties is to study neutrino-electron elastic scattering in a detector. In particular, it is more effective to probe the neutrino magnetic moment at low values of electron recoil energy. Experiments with low energy thresholds and good energy resolution are well suited for this purpose. Solar experiments such as BOREXINO \cite{XTN1}, Super-Kamiokande \cite{XTN2} and reactor experiments like GEMMA
	\cite{XTN3}, TEXONO \cite{XTN4}, MUNU \cite{XTN5} are providing some competing bounds on neutrino magnetic
	moments. However, more stringent constraint comes from astrophysical sources, such
	as globular clusters and white dwarfs \cite{XTN6}, and the recent
	XENONnT experiment also provides strict limits \cite{XTN7,XTN8}. In recent work \cite{xin1,XTN8}, they made the study in the radiative type II and type III seesaw scenario to realize neutrino electromagnetic vertex at one-loop with dark matter.
	
	Renormalization is carried out to remove the ultraviolet divergence that appears in the loop calculations. The mass-on-shell subtraction scheme is often used in the electroweak process
	calculation. The advantage of the on-shell scheme is that all parameters have a clear physical meaning and can be measured directly in the experiment \cite{onel1,onel2,yao}. The neutrino masses of the tree-level are given by the type I+II seesaw mechanism. Further, we use on-shell scheme to consider the effect of one-loop results on the neutrino mass of the tree-level. Combining the results of tree-level and one-loop, the tiny neutrino masses are given.
	Applying the effective Lagrangian method and the on-shell scheme, we analyze the radiative contributions from the one-loop diagrams to the neutrinos transition magnetic moment in the TNMSSM. In the numerical analysis, we take into the measured results in Eq.~(\ref{data}). It can be noted that the hierarchy of neutrino masses has not been determined experimentally, both the cases of NH neutrino masses and IH neutrino masses are considered in this work.

	This paper is organized as follows: The framework of the TNMSSM and the mechanism giving tiny neutrino masses are present in Sec. \ref{sec2}. The transition magnetic moment of Majorana neutrinos in the TNMSSM is calculated in Sec. \ref{sec3}. Sec. \ref{sec4} presents the numerical analyses of the neutrino transition in the TNMSSM. Conclusions are given in Sec. \ref{sec5}.
	\section{The TNMSSM}
	\label{sec2}
	Besides the superfield of the MSSM, TNMSSM introduces a guage singlet superfield $S$ and two $SU(2)_L$ triplet superfields $T$ and $\bar{T}$. The corresponding superpotential of the TNMSSM is given by \cite{TNM p1}
	\begin{eqnarray}
		W &=&S \left( \lambda H_u\cdot H_d + \lambda_T \textnormal{tr}(\bar{T} T) \right)
		+ \frac{ \kappa }{3} S^3 + \chi_u H_u \cdot \bar{T} H_u + \chi_d H_d \cdot T H_d \nonumber\\
		&&+h_u H_u\cdot Q\bar{u}+h_d H_d\cdot Q\bar{d}+h_eH_d\cdot L \bar{e}\, , \label{eq:model}
	\end{eqnarray}
where $ H_d^T = \Big( { H_d^0, H_d^ - } \Big)$, $ H_u^T = \Big( { H_u^ + , H_u^0} \Big)$, $ Q_i^T = \Big( {{{ u}_i},{{ d}_i}} \Big)$, $ L_i^T = \Big( {{{ \nu}_i},{{ e}_i}} \Big)$ are $SU(2)$ doublet superfields, and $ d_i^c$, $ u_i^c$ and $ e_i^c$ represent the singlet down-type quark, up-type quark
and charged lepton superfields, respectively. In addition $\lambda$, $\lambda_T$, $\kappa$, $\chi_u$, $\chi_d$, $h_{u,d,e}$ are dimensionless couplings. 

	Here the triplet superfields with hypercharge $Y=\pm1$ are defined as follows:
\begin{eqnarray}
	T&\equiv& T^a \sigma^a =
	\begin{pmatrix}
		T^+/\sqrt{2}&-T^{++}\\
		T^0&-T^+/\sqrt{2}
	\end{pmatrix}, \\
	\bar{T}&\equiv& \bar{T}^a\sigma^a =
	\begin{pmatrix}
		\bar{T}^-/\sqrt{2}&-\bar{T}^{0}\\
		\bar{T}^{--}&-\bar{T}^-/\sqrt{2}
	\end{pmatrix}.
\end{eqnarray}
The $\sigma^a\;(a=1,\;2,\;3)$  are $2\times2$ Pauli matrices, other products between $SU(2)_L$ doublets and $SU(2)_L$ triplets have the following form:
\begin{eqnarray}
	H_u \cdot H_d &=& H_u^+H_d^--H_u^0H_d^0 ,\\
	H_u \cdot \bar{T} H_u &=& \sqrt{2}H_u^+H_u^0\bar{T}^--\left(H_u^0\right)^2\bar{T}^0-\left(H_u^+\right)^2\bar{T}^{--}, \\
	H_d \cdot T H_d &=&\sqrt{2}H_d^-H_d^0T^+-\left(H_d^0\right)^2T^0-\left(H_d^-\right)^2T^{++}  .
\end{eqnarray}

Based on the TNMSSM superfields, we introduce three singlet right-handed neutrino superfields $N^c$ with hypercharge zero, and the superpotential involving newly introduced $N^c$ can be written as :
	\begin{eqnarray}
		W_{\textnormal{type I+II}}=Y_L L\cdot T{L}+Y_D L\cdot H_u {N^c}+Y_R N^c S {N^c},\label{eq:Type1+2}
	\end{eqnarray}

	The soft breaking term of the TNMSSM are generally given as 
	\begin{eqnarray}
		-\cal{L}_{\textnormal{soft}}
		&=& m_{H_u}^2|H_u|^2+m_{H_d}^2|H_d|^2+m_{S}^2|S|^2+m_{T}^2\textnormal{tr}(|T|^2)+ m_{\bar{T}}^2\textnormal{tr}(|\bar{T}|^2)+ +m_{\bar{\nu}}^2|\bar{\nu^c}|^2 \nonumber \\
		&&+ \; m_{Q}^2\left|Q \right|^2 +m_{\bar{u}}^2\left|\bar{u}\right|^2 +m_{\bar{d}}^2\left|\bar{d} \right|^2 +m_{L}^2\left|L \right|^2 +m_{\bar{e}}^2\left|\bar{e} \right|^2 +m_{N^c}^2|N^c|^2 \nonumber \\
		&&+ \;(A_{h_u}Q\cdot H_u\bar{u}-A_{h_d}Q\cdot H_d\bar{d}-A_{h_e}L\cdot H_d \bar{e}  + A S H_u\cdot H_d +\nonumber\\
		&&+ \; A_T S \ \textnormal{tr}(T\bar{T}) + \frac{A_{\kappa}}{3}S^3+ A_u H_u\cdot\bar{T}H_u+A_d H_d\cdot T H_d\nonumber\\
		&&+ \;A_{Y_L} L\cdot T L+A_{Y_N}  L\cdot H_u {N^c}+A_{Y_R}N^c S {N^c}+ h.c.)      \nonumber\\
		&&- \; \frac{1}{2}\Big({M_3}{{\tilde \lambda }_3}{{\tilde \lambda }_3}
		+ {M_2}{{\tilde \lambda }_2}{{\tilde \lambda }_2} + {M_1}{{\tilde \lambda }_1}{{\tilde \lambda }_1} + \textrm{h.c.} \Big)\, .\label{eq:new}
	\end{eqnarray}

	When the electroweak symmetry is broken, the neutral scalars generally gains nonzero VEVs:
	\begin{eqnarray}
		\langle H_d^0 \rangle =\frac{\upsilon_d}{\sqrt{2}} \,, \qquad \langle H_u^0 \rangle = \frac{\upsilon_u}{\sqrt{2}} \,, \qquad 	\langle T^0 \rangle =\frac{\upsilon_T}{\sqrt{2}} \,, \qquad	\langle{\bar{T^0}} \rangle =\frac{\upsilon_{\bar{T}}}{\sqrt{2}} \,, \qquad \langle S \rangle =\frac{\upsilon_S}{\sqrt{2}} \,.
	\end{eqnarray}
	Thus the neutral scalars fields can be written as usual
	\begin{eqnarray}
		&&H_d^0=\frac{h_d + i P_d + \upsilon_d}{\sqrt{2}}  , \qquad\;\bar{T}^{0}= \frac{h_{\bar{T}} + i P_{\bar{T}} + \upsilon_{\bar{T}}}{\sqrt{2}} ,  \nonumber\\
		&&H_u^0=\frac{h_u + i P_u + \upsilon_u}{\sqrt{2}}, \qquad T^0=\frac{h_T + i P_T + \upsilon_T}{\sqrt{2}}, \nonumber\\
		&&S^0=\frac{h_S + i P_S + \upsilon_S}{\sqrt{2}},
	\end{eqnarray}
	For convenience, we can define the parameters as:
	\begin{eqnarray}
	v_{ud}^{2}= v_u^2+v_{d}^2 ,\qquad\;v_{T \bar{T}}^{2}= v_T^2+v_{\bar{T}}^2 ,	\qquad\;\tan\beta=\frac{\upsilon_u}{\upsilon_d}, \qquad\; \tan\beta^{'}=\frac{\upsilon_T}{\upsilon_{\bar{T}}},\label{12}
	\end{eqnarray}

In the TNMSSM, the masses of W boson and Z boson can be written in the following form:
	\begin{eqnarray}
		M_Z^2&=&\frac{{g_1}^2+g_2^2}{4}(v_u^2+v_d^2+4v_T^2+4v_{\bar{T}}^2) , \nonumber\\ 
		M_W^2&=&\frac{g_2^2}{4}(v_u^2+v_d^2+2v_T^2+2v_{\bar{T}}^2)       ,   \nonumber\\ 
		v^2&=& v_u^2+v_d^2+2v_T^2+2v_{\bar{T}}^2 \approx (246 \textnormal{GeV})^2,\label{vev}
	\end{eqnarray}
where $g_1$ and $g_2$ denote $U(1)_Y$ and $SU(2)_L$ gauge coupling constants, respectively. 

The effective $\mu$ term is generated spontaneously via the nonzero VEVs of singlet $S$, when the electroweak symmetry is broken (EWSB)
	\begin{eqnarray}
		\mu^{\textnormal{eff}} = \lambda v_s , \ \ \ 
		\mu^{\textnormal{eff}}_T = \lambda_T v_s. 
		\label{uuuu}
	\end{eqnarray}
	
	In general for the tiny neutrino masses and mixings, the type-I seesaw  mechanism \cite{type1 p1} is the simplest and which can be realized by introducing three right-handed neutrinos to the TNMSSM superfield. Another interesting approach is known as type-II seesaw  mechanism  \cite{type2 p1}, which is realized by means of the Higgs triplet T in the TNMSSM.
	
	Then the neutrino mass term in TNMSSM can be written as
	\begin{eqnarray}
		&&{-\cal{L}}_{M_{\nu}}=\frac{1}{2}{\bar \nu}_L M_{L} {\nu}_L^C+\frac{1}{2}{\bar N}_R^C M_R {N}_R +{\bar \nu}_L M_D {N}_R+h.c.\nonumber\\
		&&\qquad\quad=\frac{1}{2} \left(\begin{array}{cc}{\bar \nu}_L &{\bar N}_R^C\end{array}\right)\left(\begin{array}{cc} M_{L}&M_D\\ M_D^T &M_R \end{array}\right)\left(\begin{array}{cc}{\nu}_L^C\\ {N}_R\end{array}\right)+h.c. \label{Mtr}
	\end{eqnarray}
with
	\begin{eqnarray}
	M_{L} =  \sqrt{2}Y_L v_T , \ \ \ 
	M_D = \frac{Y_D v_u}{\sqrt{2}} ,\ \ \ 
	M_R =  \sqrt{2}Y_R v_S ,\label{MLR}
\end{eqnarray}
where $M_D$ is the $3$ by $3$ Dirac mass term, $M_L$ and $M_R$ are the $3$ by $3$ Majorana mass terms. With the rotation matrix $Z_{N_\nu}$, the masses of neutrinos are gotten by the formula
$Z_{N_\nu}^TM_{\nu}Z_{N_\nu}=diag(m_{\nu i}), ~ i=1\dots 6$ \cite{Zhao1}. The matrix $Z_{N_\nu}$ is defined by the leading order of $\varsigma$, which is  defined as $\varsigma=M^{}_{\rm D} M^{-1}_{\rm R}$.
It is a good approximation to adopt $Z_{N_\nu}^T$ in the following form\cite{munuSSMOL}
\begin{eqnarray}
	Z_{N_\nu}^T=\left(\begin{array}{cc}
		\mathcal{S}^T&0 \\
		0 &  \mathcal{R}^T
	\end{array}\right).
	\left(\begin{array}{cc}
		1-\frac{1}{2}\varsigma^\dag\varsigma & -\varsigma^\dag\\
		\varsigma &  1-\frac{1}{2}\varsigma\varsigma^\dag
	\end{array}\right).\label{SR}
\end{eqnarray}
Here the matrices $\mathcal{S}$ and $\mathcal{R}$ defined in Eq. (\ref{SR})
are to diagonalize $M_{\nu}^\text{tree}$ and $M_R$.
\begin{eqnarray}
	&&\mathcal{S}^TM_{\nu}^\text{{tree}}\mathcal{S}=diag(m_{\nu_1},m_{\nu_2},m_{\nu_3}),\nonumber\\&&
	\mathcal{R}^TM_R
	\mathcal{R}=diag(m_{\nu_4},m_{\nu_5},m_{\nu_6}).
\end{eqnarray}

	In this condition, the effective light neutrino mass matrix of the tree-level is generally given as \cite{type12 p1,type12-p4,type12-p5,UPMNS p1,UPMNS p2,type12-p2,type12-p3}
	\begin{eqnarray}
		&&M_{\nu}^\text{{tree}} \approx m_{\nu} ^{II}+m_{\nu} ^{I}=	M_{L} - M_D M_R^{-1} M_D^{T}, \label{meff1}
	\end{eqnarray}
	where the first term  $m_{\nu} ^{II}$ is the result of type-II seesaw mechanism and the second term $m_{\nu} ^{I}$ is the result of the type-I seesaw mechanism by using the approximation $M_R^{-1}M_D^*M_D^T\approx M_DM_D^\dagger M_R^{-1}\ll M_R$. 

When the tree-level neutrino mass is obtained, we further consider the effect of one-loop radiation correction on the neutrino mass and use on-shell renormalization scheme to remove the UV divergence\cite{onel2,yao}. It can be written as
\begin{eqnarray}
	&& \delta m_{\nu{ij}}^{1-\text{loop}}=\delta m_{\nu{ij}}^{(Z,{\nu^0})}+\delta m_{\nu{ij}}^{(W,e^-)}
	+\delta m_{\nu{ij}}^{(H^0,\nu^0)}+\delta m_{\nu{ij}}^{(A^0,\nu^0)} \nonumber\\
	&&\quad \quad+\delta m_{\nu{ij}}^{(\tilde{\nu}^I,\tilde{\chi})}+\delta m_{\nu{ij}}^{(\tilde{\nu}^R,\tilde{\chi})}+\delta m_{\nu{ij}}^{(H^+,e^-)}+\delta m_{\nu{ij}}^{(\tilde{e},\chi^-)}.
	\label{ONE}
\end{eqnarray}
Here $\delta m_{\nu{ij}}^{1-\text{loop}}$ represents the one-loop radiation correction of the neutrino mass, for which detailed derivation can be found in Appendix \ref{oloop}.

Considering those one-loop corrections, the mass matrix in Eq.~(\ref{Mtr}) is rewritten as
\begin{eqnarray}
&&M_{\nu}^\text{{sum}}=M_{\nu}+Z_{N_\nu}\delta m_{\nu} Z_{N_\nu}^T\nonumber\\&&
\quad \quad \quad=\left(\begin{array}{cc}
	(M_L+\delta(m_L))_{3\times3} & (M_D+\delta(m_D))_{3\times3}\\
	(M_D+\delta(m_D))^T_{3\times3} & (M_R+\delta({m_R}))_{3\times3}

\end{array}\right).
	\label{NUMOL}
\end{eqnarray}
the matrix $M_{\nu}^\text{{sum}}$ in Eq.~(\ref{NUMOL}) including the one loop corrections also has a seesaw structure.
Similar as Eq.~(\ref{meff1}), at one-loop level we obtain the
corrected effective light neutrino mass matrix in the following form \cite{munuSSMOL}
\begin{eqnarray}
	&&{M_{\nu}^{eff}}\approx{(M_L+\delta(m_L))}-(M_D+\delta(m_D))\cdot{(M_R+\delta({m_R})))}^{-1}\cdot{(M_D+\delta(m_D))^T}.\nonumber\\
	\label{meff}
\end{eqnarray}
With the "top-down" method \cite{top-down,neu-b5,Yan1} shown in Appendix \ref{TDM}, we can diagonalize the effective neutrino mass matrix ${M_{\nu}^{eff}}$ and obtain three light neutrino masses, mixing angles with neutrinos.

In the
leading-order approximation, the effective mass matrix of three
light neutrinos is given by $M_{\nu}^{eff} \approx (M_L+\delta(m_L)) -
(M_D+\delta(m_D))\cdot{(M_R+\delta({m_R})))}^{-1}\cdot{(M_D+\delta(m_D))}$. Either $M_D+\delta(m_D)$ or
$(M_D+\delta(m_D))\cdot{(M_R+\delta({m_R})))}^{-1}\cdot{(M_D+\delta(m_D))}$ may dominate $M_{\nu}^{eff}$,
 but another possibility can be focused on: the smallness of
$M_{\nu}^{eff}$ arises from a significant cancellation between $M_L+\delta(m_L)$ and $(M_D+\delta(m_D))\cdot{(M_R+\delta({m_R})))}^{-1}\cdot{(M_D+\delta(m_D))}$ in the case of
${\cal O}(M_{\nu}^{eff}) \ll {\cal O} (M_L+\delta(m_L)) \sim {\cal
	O}((M_D+\delta(m_D))\cdot{(M_R+\delta({m_R}))}^{-1}\cdot{(M_D+\delta(m_D))})$. The tiny neutrino masses implies that the relation
${\cal O} (M_L+\delta(m_L)) \sim {\cal
	O}((M_D+\delta(m_D))\cdot{(M_R+\delta({m_R}))}^{-1}\cdot{(M_D+\delta(m_D))})$ must hold. It is the significant but
incomplete cancellation between $M_L+\delta(m_L)$ and $(M_D+\delta(m_D))\cdot{(M_R+\delta({m_R}))}^{-1}\cdot{(M_D+\delta(m_D))}$ terms that results in the non-vanishing
but tiny masses for three light neutrinos \cite{CW1,CW2,CW3}. In this interesting case, it will enhance the $Y_L$ and $Y_D$ terms in Eq.~(\ref{MLR}), which affects the coupling of the neutrino to fermions and scalars. Therefore, this will have an impact on the transition magnetic moment and neutrino mass of the Majorana neutrinos that are studied in this work. According to the above analysis, we make $M_{\nu}^\text{{tree}}$ to give the tree-level neutrino mass and combine the results given by the tree-level and the one-loop corrections to give the mass of the neutrino that satisfies the strict neutrino experimental limit. Specific details will be discussed in the numerical analysis \ref{sec4}. 
	
	\section{Neutrino magnetic moment}
	\label{sec3}
	The electric dipole moment (EDM)  and  magnetic dipole moment (MDM) of the Dirac
	fermion (e.g. charged lepton, neutrino, etc) can be written as the operators
	\begin{eqnarray}
		&&\:\mathcal{L}_{EDM}=\frac{i}{2} \epsilon_{ij} \bar{\psi}_i \sigma^{\mu\nu} \gamma_5 \psi_j F_{\mu\nu},
		\label{MEDM}\nonumber\\
		&&\:\mathcal{L}_{MDM}=\frac{1}{2} \mu_{ij} \bar{\psi}_i \sigma^{\mu\nu} \psi_j F_{\mu\nu},
	\end{eqnarray}
	where $\sigma^{\mu\nu}=\frac{i}{2}[\gamma^\mu,\gamma^\nu]$, $F_{\mu\nu}$ is the electromagnetic field strength, $\psi_{i,j}$ is the four component Dirac fermions of on-shell,  $\epsilon_{ij}$ and $\mu_{ij}$ are Dirac diagonal ($i=j$) or transition
	($i\neq j$) EDM and MDM between states $\psi_{i}$ and $\psi_{j}$, respectively.
	
To obtain the Dirac fermion EDM and MDM, we use the effective Lagrangian method. The reason is that the masses of internal lines are much greater than the masses of external Dirac fermion in the TNMSSM, then it is more convenient to employ the effective Lagrangian method to calculate the contributions from loop diagrams to fermion diagonal or transition EDM and MDM \cite{EFT1}. It is sufficient to retain only the dimensional-6 operators in the later calculations \cite{EFT2,Feng,Feng1,Feng2}:
	\begin{eqnarray}
		&&O_1^{L,R} = e \bar{\psi}_i {(i {/\!\!\!\! \mathcal{D}})}^3 P_{L,R} \psi_j , \nonumber\\
		&&O_2^{L,R} = e \overline{(i \mathcal{D}_\mu {\psi}_i )} \gamma^\mu F\cdot \sigma P_{L,R} \psi_j, \nonumber\\
		&&O_3^{L,R} = e \bar{\psi}_i  F\cdot \sigma \gamma^\mu P_{L,R} {(i \mathcal{D}_\mu {\psi}_j )}, \nonumber\\
		&&O_4^{L,R} = e \bar{\psi}_i  (\partial^\mu F_{\mu\nu})  \gamma^\nu P_{L,R} \psi_j, \nonumber\\
		&&O_5^{L,R} = e m_{{\psi}_i} \bar{\psi}_i  {(i {/\!\!\!\! \mathcal{D}})}^2 P_{L,R} \psi_j, \nonumber\\
		&&O_6^{L,R} = e m_{{\psi}_i} \bar{\psi}_i F\cdot \sigma P_{L,R} \psi_j,
		\label{operators}
	\end{eqnarray}
	where $\mathcal{D}_\mu=\partial^\mu+ieA_\mu$,  $P_L=\frac{1}{2}{(1 - {\gamma _5})}$, $P_R=\frac{1}{2}{(1 + {\gamma _5})}$ and $m_{{\psi}_i}$ is the mass of fermion ${{\psi}_i}$.
	
	\begin{figure}
		\setlength{\unitlength}{1mm}
		\centering
		\includegraphics[width=4.5in]{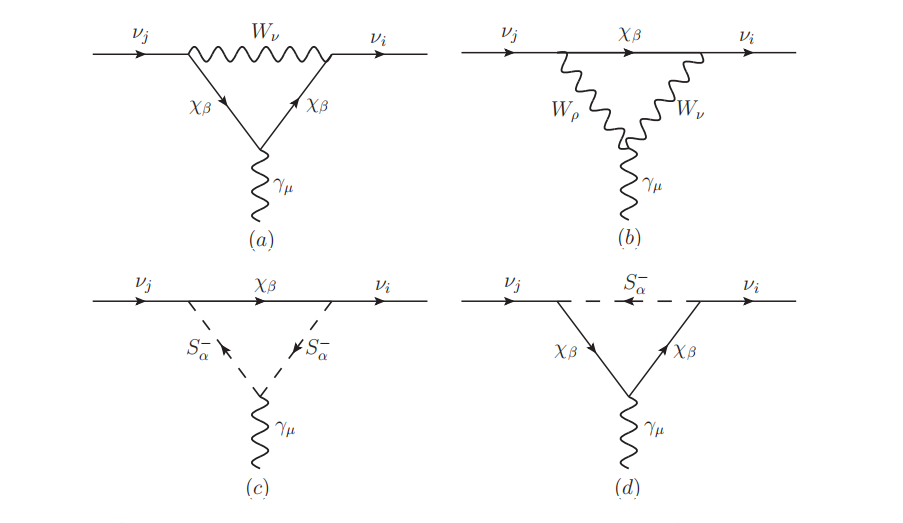}
		\vspace{0cm}
		\caption[]{One-loop diagrams contributing to transition magnetic moment of Majorana neutrinos in the TNMSSM, where (a) and (b) are the charged fermion $\chi_\beta$ and $W$-boson loop contributions, and (c) and (d) are the charged fermion $\chi_\beta$ and charged scalar $S_\alpha ^-$ loop contribution.}
		\label{feynman}
	\end{figure}

 By describing the electromagnetic form factors of Dirac and Majorana neutrinos in the Ref.~\cite{Broggini}, one can obtain the EDM and MDM for Majorana neutrinos of the following form:
	\begin{eqnarray}
		&&\epsilon_{ij}^M = \epsilon_{ij}^D -\epsilon_{ji}^D,\qquad\mu_{ij}^M = \mu_{ij}^D -\mu_{ji}^D,
		\label{Majorana-MDM}
	\end{eqnarray}
	with
	\begin{eqnarray}
		&&\:\epsilon_{ij}^D = 4m_e m_{\nu_i} \Im(C_2^{R} + \frac{m_{\nu_j}}{m_{\nu_i}}C_2^{L\ast} + C_6^{R}) \mu_{\rm{B}},\nonumber\\
		&&\mu_{ij}^D = 4m_e m_{\nu_i} \Re(C_2^{R} + \frac{m_{\nu_j}}{m_{\nu_i}}C_2^{L\ast} + C_6^{R}) \mu_{\rm{B}},
	\label{28}
	\end{eqnarray}
	$\nu_{i}$ and $\nu_{j}$ denote the Majorana neutrinos. In Eq.~(\ref{Majorana-MDM}), the first term is the Dirac neutrino like term representing the EDM and MDM, and the second term   $-\mu_{ji}^D$ and $-\epsilon_{ji}^D$ represents the Dirac anti-neutrino like term for the EDM and MDM, and the details of their derivation can be found in Appendix \ref{MDM}. It can be seen to observe that $\mu_{ij}^M$ and $\epsilon_{ij}^M$ are antisymmetric, so the EDM and MDM of the Majorana neutrinos will not be diagonal, but there can have transition EDM and MDM.
	
	In the TNMSSM, the one-loop diagrams contributing to the transition magnetic moment of the Majorana neutrinos are depicted the  Fig.~\ref{feynman}. The corresponding Wilson coefficients can be written as
	\begin{eqnarray}
		C_{2,6}^{L,R}=C_{2,6}^{{L,R}(a)}+C_{2,6}^{{L,R}(b)}+C_{2,6}^{{L,R}(c)}+C_{2,6}^{{L,R}(d)}.
	\end{eqnarray}
	Here the term  $C_{2,6}^{{L,R}(a,b)}$  represent the loop contributions from Fig.~\ref{feynman}(a) and Fig.~\ref{feynman}(b):
	\begin{eqnarray}
		&&C_2^{R(a)}= \frac{1}{2{m_W^2}} C_R^{W {\chi _\beta} \bar \nu _{i}  } C_R^{W \nu _{j} {{\bar \chi }_\beta}} \Big[  {I_1}({x_{\chi _\beta },{x_{W }}}) - {I_4}({x_{\chi _\beta },{x_{W }}}) \Big], \nonumber\\
		&&C_6^{R(a)}=  \frac{2{m_{\chi _\beta }}}{{m_W^2}{m_{\nu_i} }}  C_L^{W {\chi _\beta} \bar \nu _{i} } C_R^{W\nu _{j} ^ \circ {{\bar \chi }_\beta}} \Big[ {I_3}({x_{\chi _\beta },{x_{W }}}) -  {I_1}({x_{\chi _\beta },{x_{W }}}) \Big],  \nonumber\\
		&&C_2^{R(b)}=  \frac{1}{2{m_W^2}} C_R^{W {\chi _\beta} \bar \nu _{i} } C_R^{W \nu _{j} {{\bar \chi }_\beta}} \Big[  {I_3}({x_{\chi _\beta },{x_{W }}}) +  {I_4}({x_{\chi _\beta },{x_{W }}})\Big], \nonumber\\
		&&C_6^{R(b)}=  \frac{2{m_{\chi _\beta }}}{{m_W^2}{m_{\nu_i} }}   C_L^{W {\chi _\beta} \bar \nu _{i} } C_R^{W \nu _{j} {{\bar \chi }_\beta}} \Big[  -  {I_3}({x_{\chi _\beta },{x_{W }}})  \Big], \nonumber\\
		&&C_{2,6}^{L(a,b)}=C_{2,6}^{R(a,b)}\mid _{L \leftrightarrow R},
	\end{eqnarray}
	The concrete expressions  $I_k\:(k=1,\cdots,4)$ can be found in Ref.\cite{Zhang1,yang}, and  $x_i = {m_i^2}/{m_W^2}$ with $m_i $ denoting the mass of the corresponding particle.
	
	The loop contributions from Fig.~\ref{feynman}(c) and Fig.~\ref{feynman}(d) can be written as
	\begin{eqnarray}
		&&C_2^{R(c)}=  \frac{1}{4{m_W^2}} C_R^{S_\alpha^{-\ast} {\chi _\beta} \bar  \nu _{i} } C_L^{S_\alpha^- \nu _{j} {{\bar \chi }_\beta}} \Big[ {I_4}({x_{\chi _\beta },{x_{S_\alpha^- }}}) - {I_3}({x_{\chi _\beta },{x_{S_\alpha^- }}}) \Big], \nonumber\\
		&&C_6^{R(c)}=  \frac{{m_{\chi _\beta }}}{2{m_W^2}{m_{\nu_i} }}  C_R^{S_\alpha^{-\ast} {\chi _\beta} \bar \nu _{i} } C_R^{S_\alpha^- \nu _{j} {{\bar \chi }_\beta}} \Big[ {I_3}({x_{\chi _\beta },{x_{S_\alpha^- }}}) - {I_1}({x_{\chi _\beta },{x_{S_\alpha^- }}})  \Big],  \nonumber\\
		&&C_2^{R(d)}= \frac{1}{4{m_W^2}} C_R^{S_\alpha^{-\ast} {\chi _\beta} \bar  \nu _{i} } C_L^{S_\alpha^-  \nu _{j} {{\bar \chi }_\beta}} \Big[ 2{I_3}({x_{\chi _\beta },{x_{S_\alpha^- }}}) - {I_1}({x_{\chi _\beta },{x_{S_\alpha^- }}}) - {I_4}({x_{\chi _\beta },{x_{S_\alpha^- }}}) \Big], \nonumber\\
		&&C_6^{R(d)}= \frac{{m_{\chi _\beta }}}{2{m_W^2}{m_{\nu_i} }}   C_R^{S_\alpha^{-\ast} {\chi _\beta} \bar  \nu _{i} } C_R^{S_\alpha^-  \nu _{j} {{\bar \chi }_\beta}} \Big[{I_1}({x_{\chi _\beta },{x_{S_\alpha^- }}})-{I_2}({x_{\chi _\beta },{x_{S_\alpha^-}}}) -   {I_3}({x_{\chi _\beta },{x_{S_\alpha^- }}}) \Big],  \nonumber\\
		&& C_{2,6}^{L(c,d)}=C_{2,6}^{R(c,d)}\mid _{L \leftrightarrow R}.\label{C6R}
	\end{eqnarray}

	\section{The numerical analyses}
	\label{sec4}
	In the calculation, we take the $W$ boson mass $m_W=80.377\;{\rm GeV}$, the $Z$ boson mass $m_Z=90.188\;{\rm GeV}$, the electron mass $m_e=0.511\;{\rm MeV}$, $\alpha_{em}(m_Z)=1/128.9$  for the coupling of the electromagnetic
	interaction, $\alpha_s(m_Z)=0.118$ for the coupling of the strong interaction.  The constraint on the sum of neutrino masses $\sum_i m_{\nu i}<0.12\;{\rm eV}$ is considered \cite{PDG,NMTM}. So far the neutrino mass spectrum is not fixed, both the NH neutrino masses $m_{\nu1}<m_{\nu2}<m_{\nu3}$ and the IH neutrino masses $m_{\nu3}<m_{\nu1}<m_{\nu2}$ are considered in the following analyses.

The measured mass of the Higgs boson is  \cite{PDG}
	\begin{eqnarray}
		&&m_h=125.25\pm0.17\;{\rm GeV}.
		\label{higgs ma}
	\end{eqnarray}

For simplicity, we appropriately set $M_{1}=500\; {\rm GeV}$, for those Higgsino parts. Limited on supersymmetric particle masses from the Particle Data Group~\cite{PDG}, we choose  $M_2\geq300\:{\rm{GeV}}$ in the numerical calculations. Assuming that the mass parameter of slepton can be written as  $m_{\tilde L}=m_{\tilde e}=m_{\tilde \nu}={\rm diag}(M_E,M_E,M_E)\;{\rm TeV}$. The LHC experimentally excludes the case where the mass of the slepton is less than $700\; {\rm GeV}$, and here we make $M_E\gtrsim 0.8\;{\rm TeV}$ \cite{data1,data2,data3}. 

The neutrino oscillation experimental data \cite{PDG} and the
lightest CP-even Higgs boson mass constrain relevant parameter space strongly. 
 For convenience, we choose the relevant parameters as default values below for numerical calculations to reduce the number of free parameters in the model
 considered here,
		\begin{eqnarray}
		&&	\tan \beta=3.4 , \quad \lambda=0.5 ,  \quad \lambda_T=0.22 \quad \kappa=0.86  , \nonumber\\
		&&\chi_u=0.1, \quad\;  \chi_d=1.2, \;\quad   A_{\lambda_ T}= A_\lambda =1000\:{\rm{GeV}},       \nonumber\\
		&&A_\kappa =-1000\:{\rm{GeV}} , \quad\;  A_{\chi_u}=-850\:{\rm{GeV}} ,\quad\;A_{\chi_d}=-500\:{\rm{GeV}}.
	\end{eqnarray}
The relevant couplings $Y_{R,ll}$ and $Y_{D,ll}$ are not free parameters. When the $M_{R_{ll}} (l=1,2,3)$ is given, in combination with Eqs.~(\ref{vev},~\ref{uuuu},~\ref{MLR},~\ref{meff}), we can replace $Y_{R,ll}$ and $Y_{D,ll}$ with other relevant parameters. Therefore only the coupling $Y_L$ of the type-II seesaw mechanism needs to be set. Generally, $Y_{L}$ is assumed to be diagonal and takes the form:
\begin{eqnarray}
	Y_L={\rm diag}\;(Y_{ee},\;Y_{\mu\mu},\;Y_{\tau\tau}). \label{Y-matrix}
\end{eqnarray}
$Y_{ee}$ is constrained strongly by the $0\nu2\beta$ decay experiments in the range $Y_{ee}\lesssim0.04$. For simplicity in the subsequent analysis we set $Y_{ee}=Y_{\mu\mu}=Y_{\tau\tau}=Y_{LL}$.
In addition, a small VEV $v_T$ of $T^0$ and $v_{\bar{T}}$ of ${\bar{T}}$, are constrained by the $\rho-$parameter~\cite{PDG}, so later we will set $v_{T \bar{T}}=0.001\;{\rm GeV}$ to simplify the numerical evaluations.
\begin{figure}[h]
	\setlength{\unitlength}{1mm}
	\centering
	\includegraphics[width=2.5in]{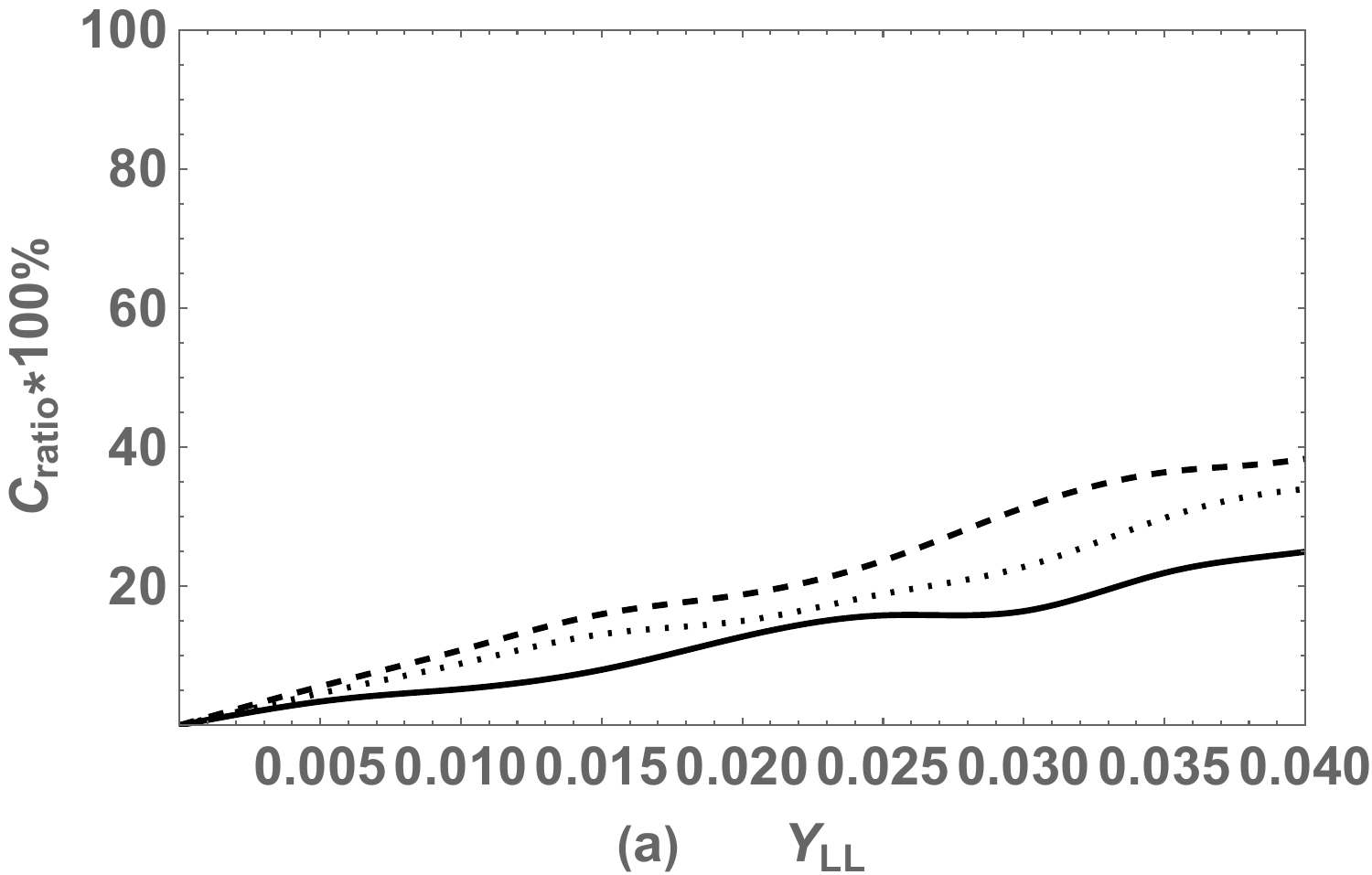}
	\vspace{0.0cm}
	\includegraphics[width=2.5in]{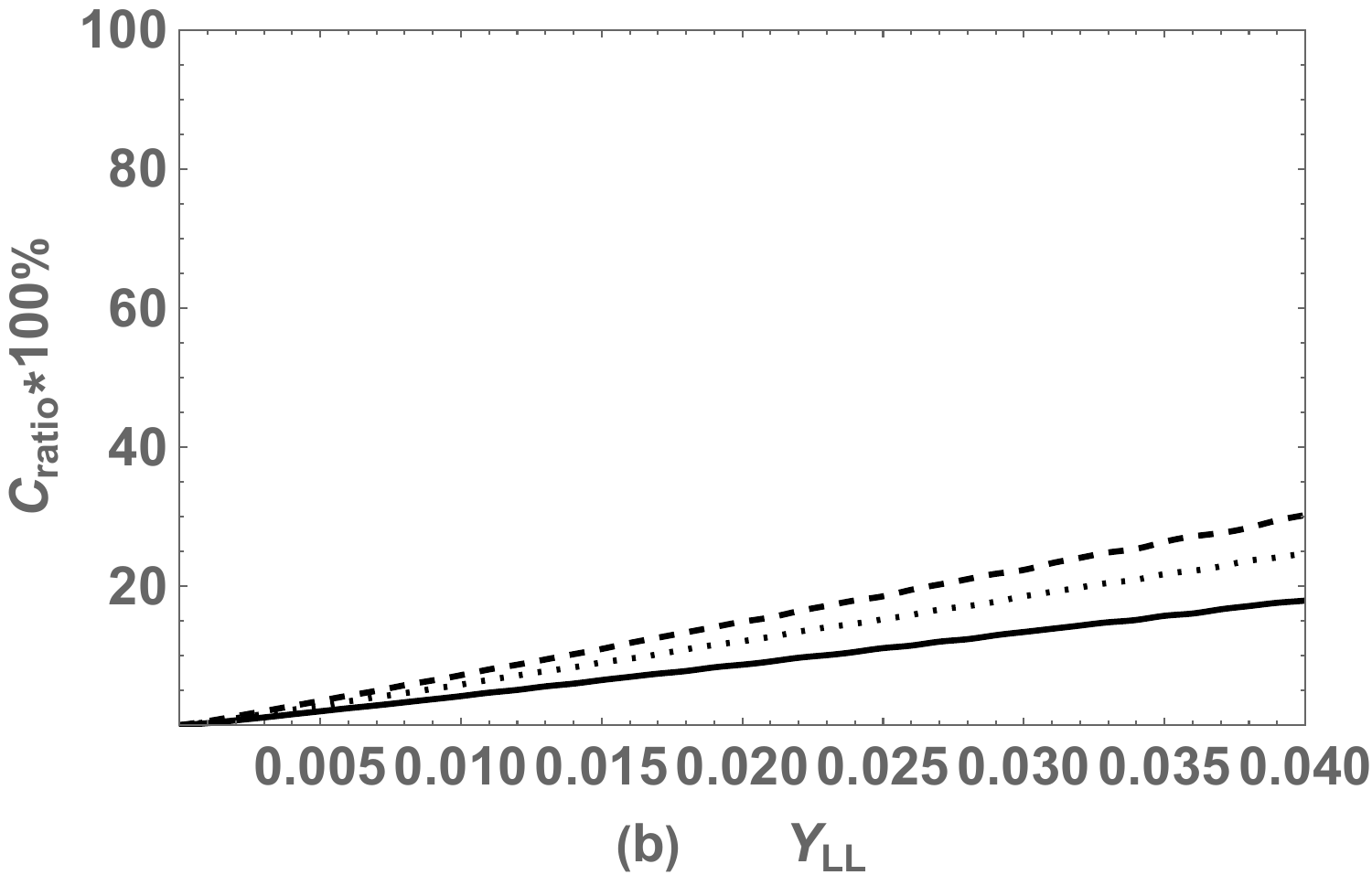}
	\vspace{0.0cm}
	\caption{$C_{\text{ratio}}$ versus $Y_{LL}$ are plotted for NH (a) and IH (b) neutrino masses, where the solid, dotted, dashed lines represent $\tan \beta^{'}=0.4$, $\tan \beta^{'}=0.6$ and $\tan \beta^{'}=0.8$ respectively.}
	\label{YLLNM}
\end{figure}

For Majorana neutrinos, their transition magnetic moments are on the order of $10^{-24}\mu_B$ \cite{nu47,nu48}. These values are much lower than the sensitivity of the present experiments. Nonstandard interactions of the neutrinos can lead to enhanced magnetic moments \cite{zy1}. In the general Standard Model extension, the tiny Majorana neutrinos masses via the type I seesaw mechanism, so $Y_{D}$ tends to be relatively small, which will depress the transition magnetic moments of Majorana neutrinos\cite{yang,fn1,Yang2}. However, in the TNMSSM, the type I+II seesaw mechanism is naturally present, which may provide an opportunity to increase the neutrino's interaction with other particles. In this way, the transition magnetic moments of Majorana neutrinos can be enhanced if the tiny neutrino mass is obtained.
According to the above analysis, the contribution of Fig.~\ref{feynman} can enhance the transition magnetic moment in the TNMSSM. The mixing parameters of the light neutrinos with charged fermions and charged scalars are not tiny under the type I+II seesaw, and the right-handed neutrinos mass matrix $M_R$ also affects the numerical results by influencing the mixing of the light neutrinos with charged fermions and charged scalars. For simplicity and not to lose general features, we assume there are no off-diagonal elements in the matrix $M_R$ and the diagonal elements are all degenerate, which means $ M_{R,11}=M_{R,22}=M_{R,33}=M_{R}$
in Eq.~(\ref{MLR}). In the later analysis, we will set $M_{R}=100\:{\rm{GeV}}$ and combine the tree-level with the one-loop results to give the neutrino mass that satisfies the neutrino oscillation experiment. The effect of one-loop corrections on neutrino mass is analyzed by the following formula
\begin{eqnarray}
	C_{\text{ratio}}=\frac{(m_{\nu-{\rm lightest}}-m_{\nu-{\rm lightest}}^\text{{tree}})}{m_{\nu-{\rm lightest}}}. \label{CR}
\end{eqnarray} 
Here $m_{\nu-{\rm lightest}}$ is the lightest neutrino mass of the NH and IH neutrino masses, and $m_{\nu-{\rm lightest}}^\text{{tree}}$ is the lightest neutrino mass of the tree-level. Then the effect of one-loop correction on neutrino mass can be reflected by Eq.~(\ref{CR}). We set  $m_{\nu-{\rm lightest}}=0.005\:{\rm{eV}}$ for NH and IH neutrino masses, $s_{12}^2,\,s_{13}^2,s_{23}^2,\,\Delta m_{\nu_{21}}^2,\,|\Delta m_{\nu_{32}}^2|$ with the center values. 

In Fig.~\ref{YLLNM}, we take $M_{2}=500 \:{\rm{GeV}} $,  $\mu=650 \;{\rm GeV}$ for NH (a) and IH (b) neutrino masses. Then the ratio of one-loop corrections versus
$Y_{LL}$ is plotted in Fig.~\ref{YLLNM}(a) and Fig.~\ref{YLLNM}(b), where the solid, dotted, dashed lines denote the results of $\tan \beta^{'}=0.4$, $\tan \beta^{'}=0.6$ and $\tan \beta^{'}=0.8$ respectively. With the increase of $Y_{LL}$, the proportion of one-loop corrections will also increase, and the proportion of one-loop corrections will further increase with the increase of $\tan \beta^{'}$. This is because $Y_{LL}$ can influence $Y_{D}$ through Eq.~\ref{MLR} and Eq.~\ref{meff1}, which in turn affects the interaction of neutrinos with fermions and scalar bosons. As for the contribution of the one-loop corrections, when $Y_{LL}$ increases, $Y_{D}$ should also increase accordingly, and the contribution of one-loop corrections will become more significant. In addition, when $v_{T \bar{T}}$ is fixed, $v_{T}$ will increase with $\tan \beta^{'}$, which in turn will have an impact on $Y_{D}$ through Eq.~\ref{MLR} and Eq.~\ref{meff1}, which in turn will affect the contribution of a one-loop. So we can find that with the increase of $Y_{LL}$ and $\tan \beta^{'}$, the contribution of the one-loop becomes more significant.

\begin{figure}[h]
	\setlength{\unitlength}{1mm}
	\centering
	\includegraphics[width=2.5in]{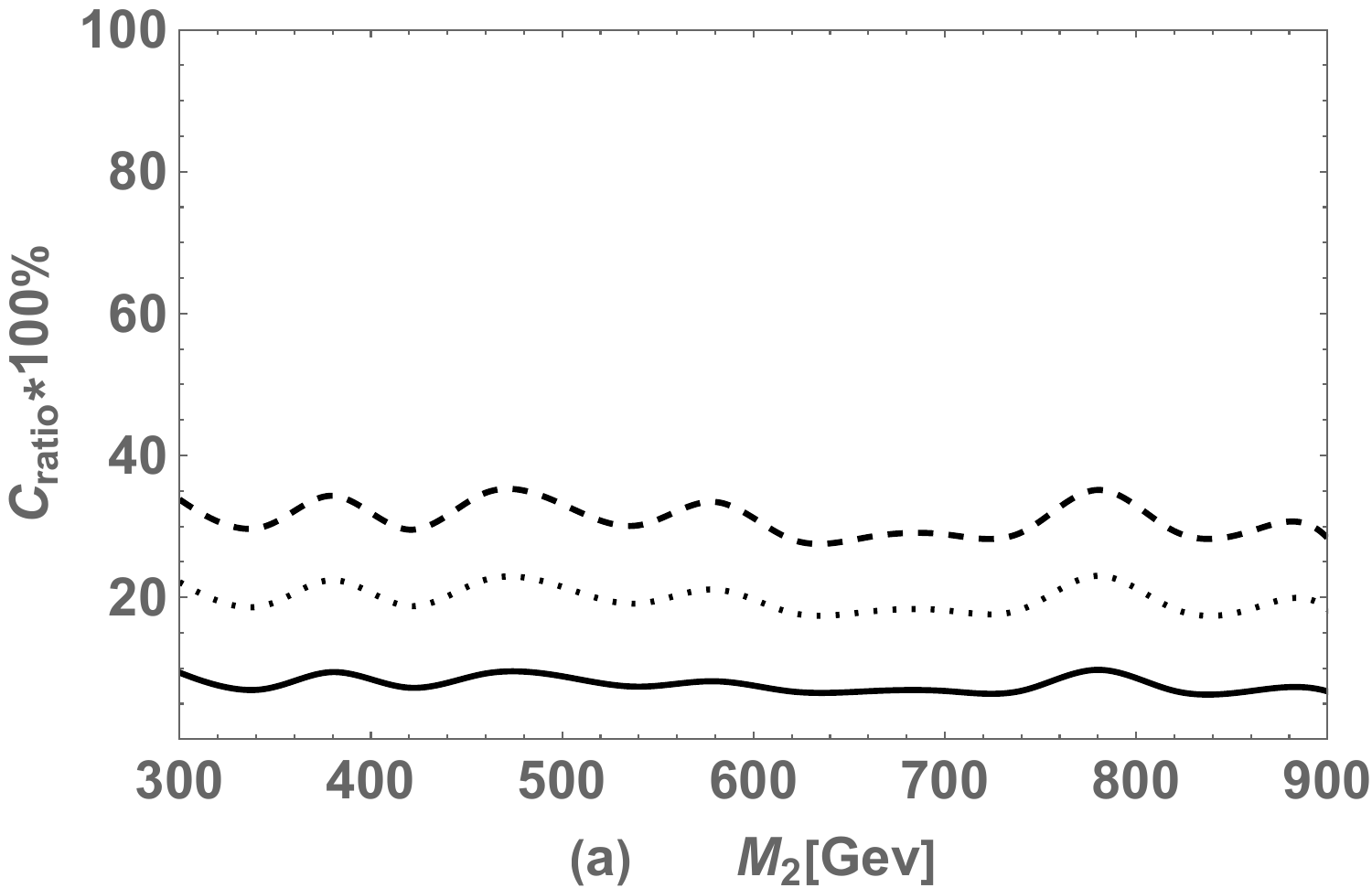}
	\vspace{0.0cm}
	\includegraphics[width=2.5in]{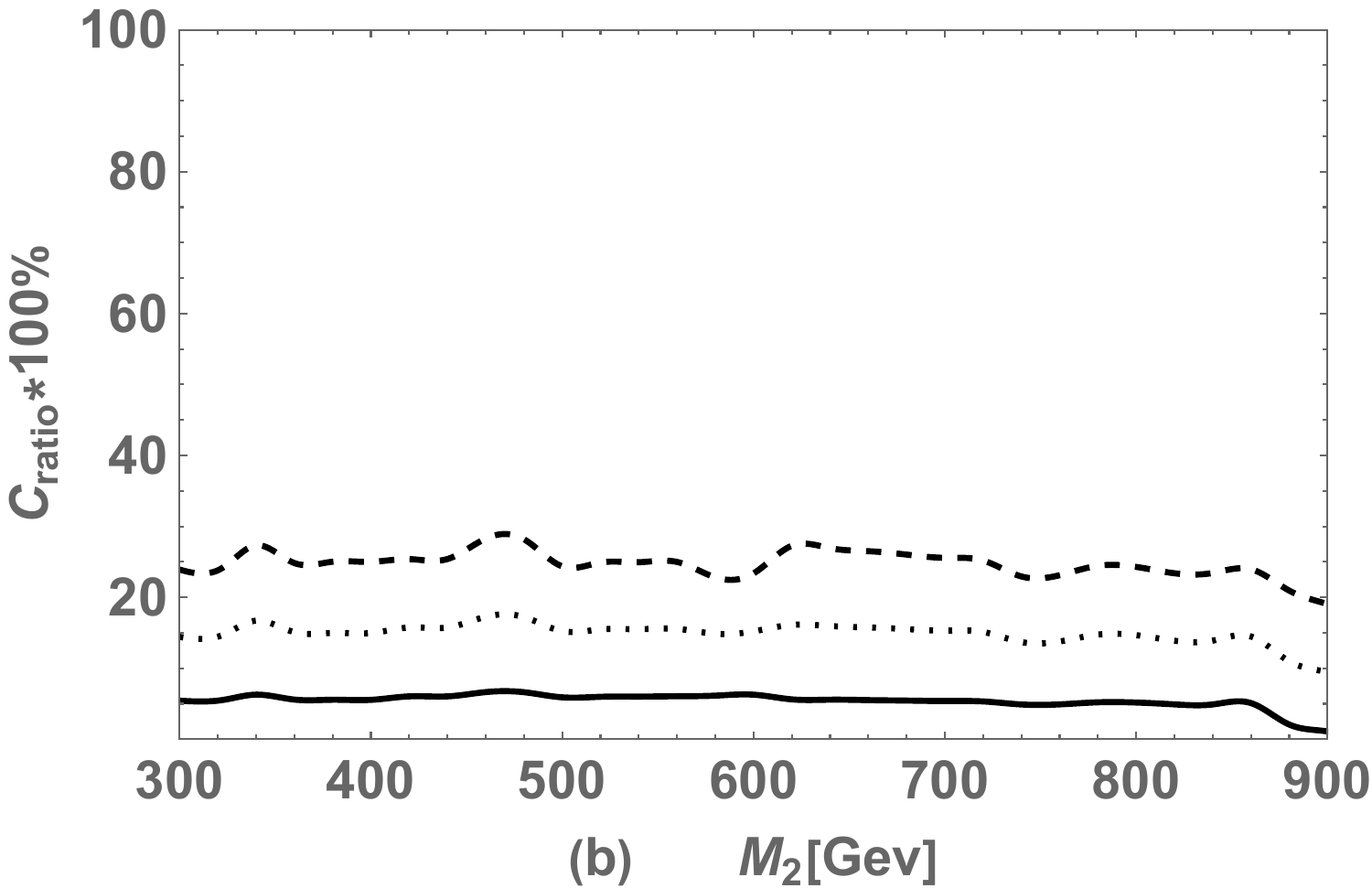}
	\vspace{0.0cm}
	\caption{$C_{\text{ratio}}$ versus $M_{2}$ are plotted for NH (a) and IH (b) neutrino masses, where the solid, dotted, dashed lines represent $Y_{LL}=0.01$, $Y_{LL}=0.025$ and $Y_{LL}=0.04$ respectively.}
	\label{M2XNM}
\end{figure}
In Fig.~\ref{M2XNM}, we take $\mu=650 \;{\rm GeV}$, $\tan \beta^{'}=0.6$ for NH (a) and IH (b) neutrino masses. Then the ratio of one-loop corrections versus
$M_{2}$ is plotted in Fig.~\ref{M2XNM}(a) and Fig.~\ref{M2XNM}(b), where the solid, dotted, dashed lines denote the results of $Y_{LL}=0.01$, $Y_{LL}=0.025$ and $Y_{LL}=0.04$ respectively. As $M_{2}$ increases, the proportion of one-loop corrections will decrease, and there will be a certain complex correlation. This relationship is further amplified as  $Y_{LL}$ increases. The soft breaking wino mass $M_2$ influences the neutralinos and chargino masses. And, the contribution of neutralinos and chargino loop diagrams is an important part of one-loop correction. As $M_2$ increases, the neutralinos and chargino mass also increase, resulting in some resonant effects from the one-loop contribution. 
Therefore, it can be seen from Fig.~\ref{M2XNM} that fluctuations occur with the change of $M_2$. This is caused by the interference effect between neutralino-sneutrino and chargino-slepton loop diagrams.

\begin{figure}[h]
	\setlength{\unitlength}{1mm}
	\centering
	\includegraphics[width=2.5in]{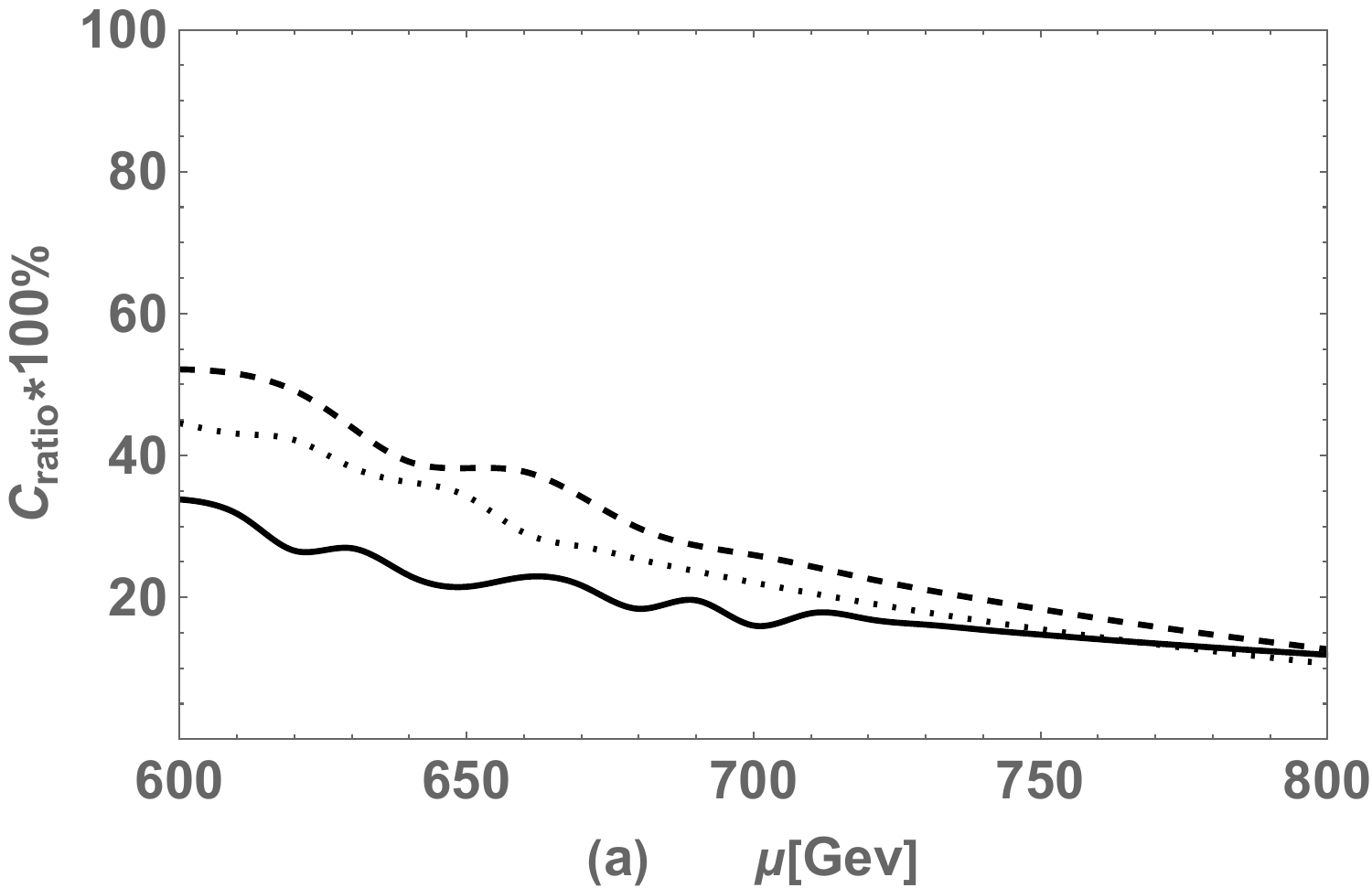}
	\vspace{0.0cm}
	\includegraphics[width=2.5in]{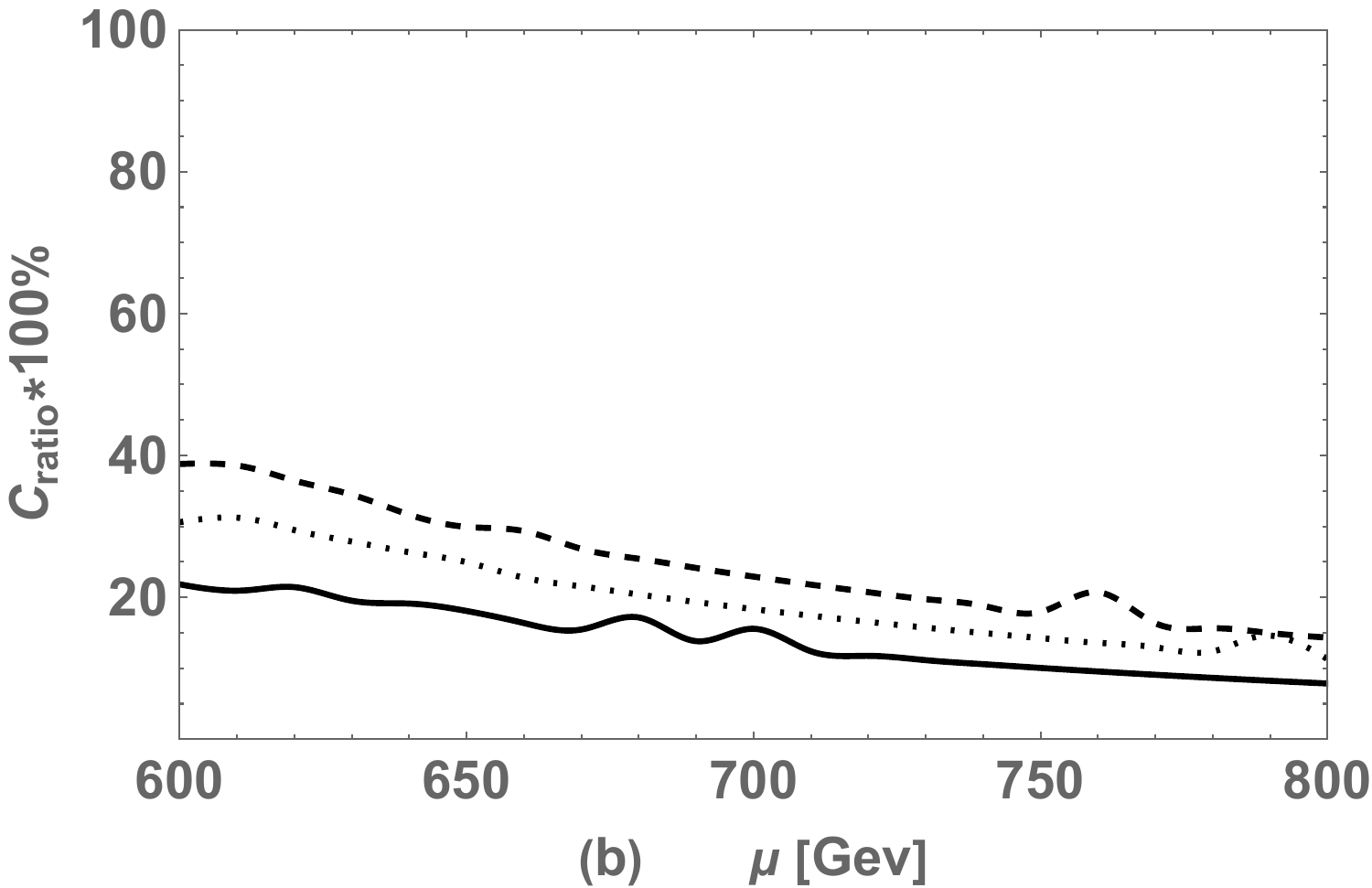}
	\vspace{0.0cm}
	\caption{$C_{\text{ratio}}$ versus $\mu$ are plotted for NH (a) and IH (b) neutrino masses, where the solid, dotted, dashed lines represent $\tan \beta^{'}=0.4$, $\tan \beta^{'}=0.6$ and $\tan \beta^{'}=0.8$ respectively.}
	\label{TBNM}
\end{figure}

In Fig.~\ref{TBNM}, we take $M_2=500 \;{\rm GeV}$, $Y_{LL}=0.04$ for NH (a) and IH (b) neutrino masses. Then the ratio of one-loop corrections versus
$\mu$ are plotted in Fig.~\ref{TBNM}(a) and Fig.~\ref{TBNM}(b), where the solid, dotted, dashed lines denote the results of $\tan \beta^{'}=0.4$, $\tan \beta^{'}=0.6$ and $\tan \beta^{'}=0.8$ respectively. With the increase of $\mu$, the proportion of one-loop corrections will decrease. And as $\tan \beta^{'}$ increases, the proportion of one-loop corrections increases further. This is because $\mu$ term can influence $v_{S}$ through Eq.~\ref{uuuu}, which in turn affects the mass of neutralinos, chargino, sleptons, and charged Higgs, but changes in $\mu$ have a greater effect on neutralino and chargino masses than other particles. As $\mu$ increases, the mass of the associated particles also increases, leading to the suppression of the sneutrino-neutralino and the slepton-charginos loop-diagram contribution. From Fig.~\ref{M2XNM} and Fig.~\ref{TBNM}, we can see that the change in the mass of the neutralinos and charginos greatly affects the proportion of the result of the one-loop, and perhaps there is a deeper connection between the neutrinos and neutralinos and charginos. 

\begin{figure}
	\setlength{\unitlength}{1mm}
	\centering
	\begin{minipage}[c]{0.5\textwidth}
		\includegraphics[width=2.9in]{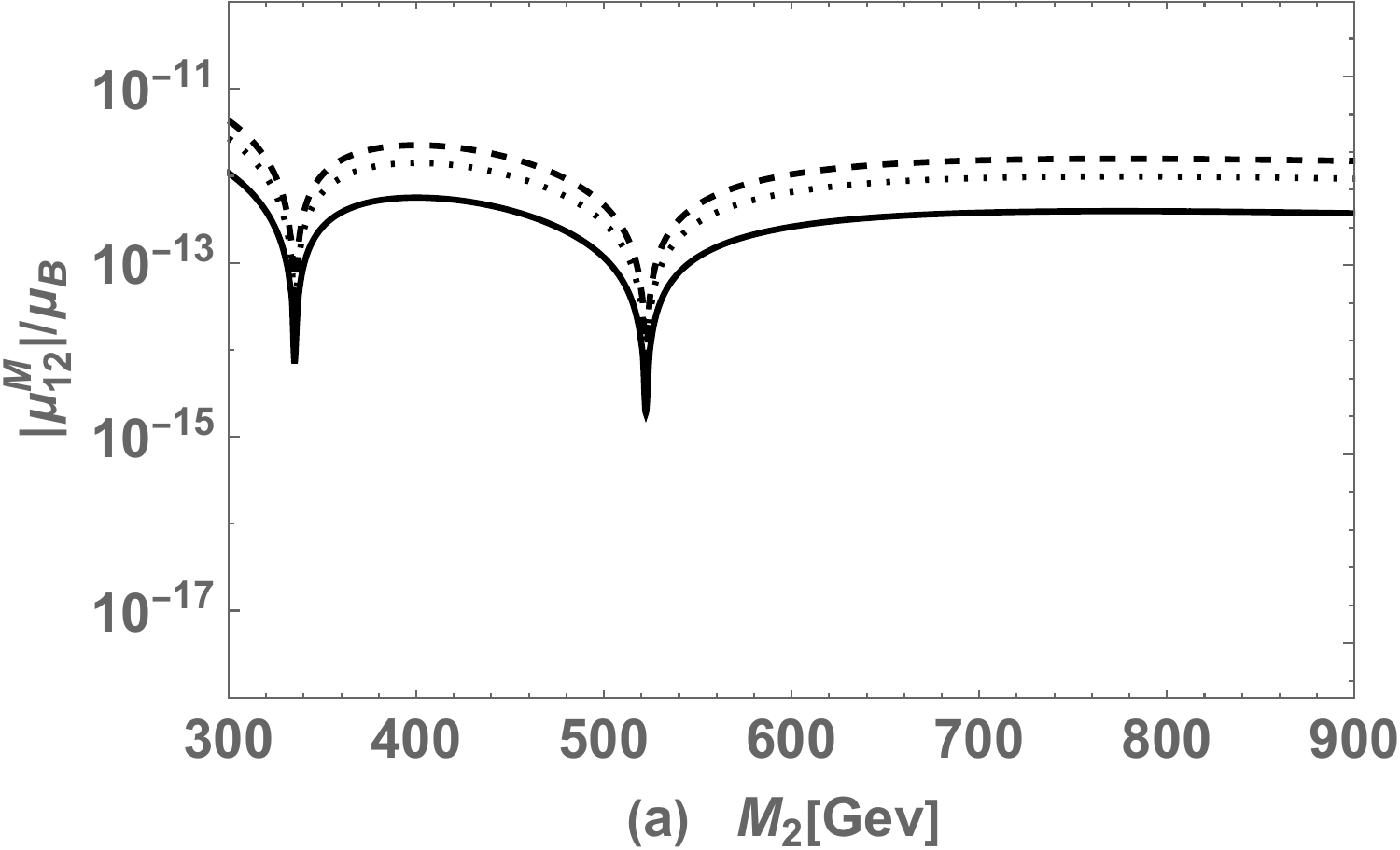}
	\end{minipage}%
	\begin{minipage}[c]{0.5\textwidth}
		\includegraphics[width=2.9in]{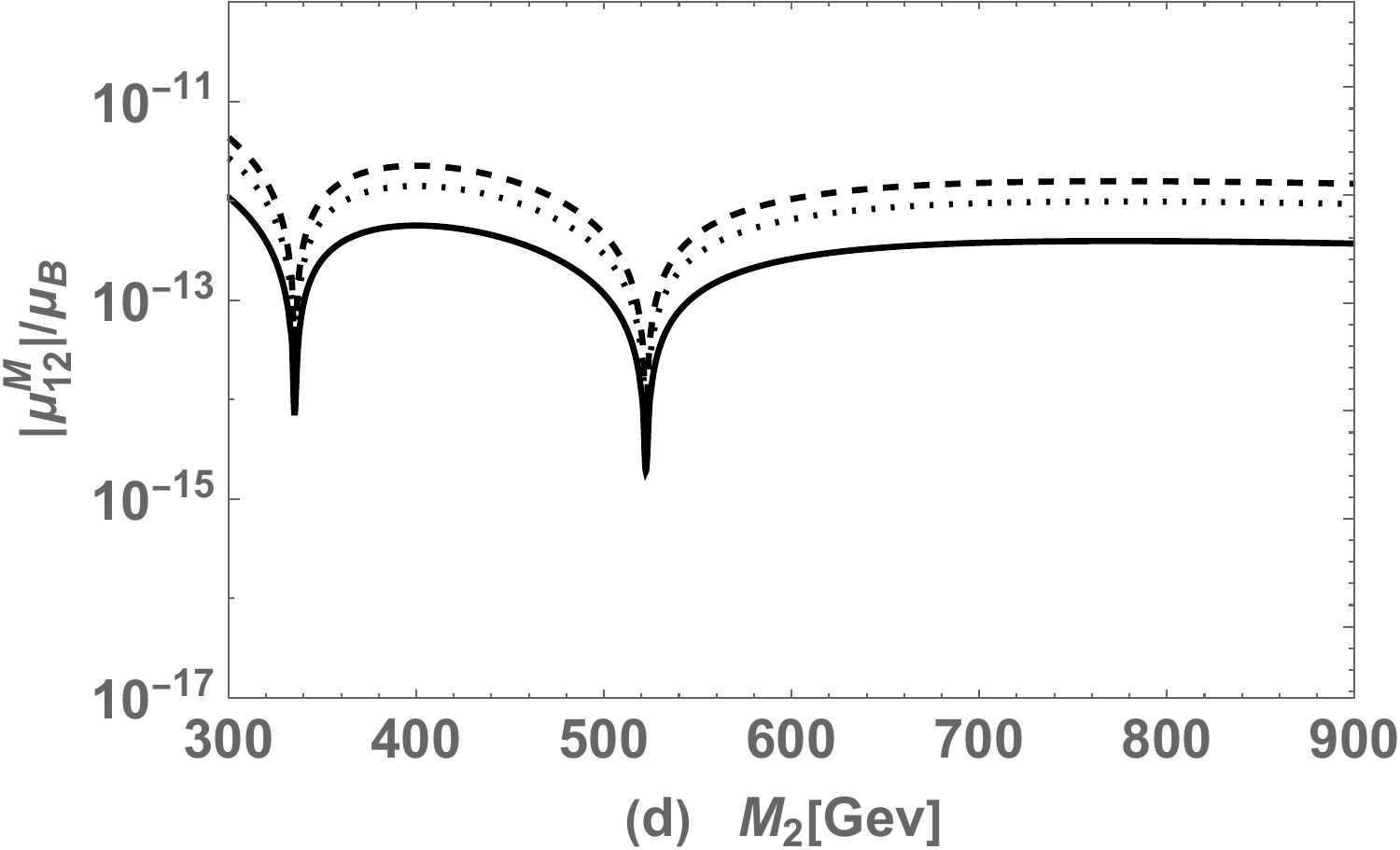}
	\end{minipage}
	\begin{minipage}[c]{0.5\textwidth}
		\includegraphics[width=2.9in]{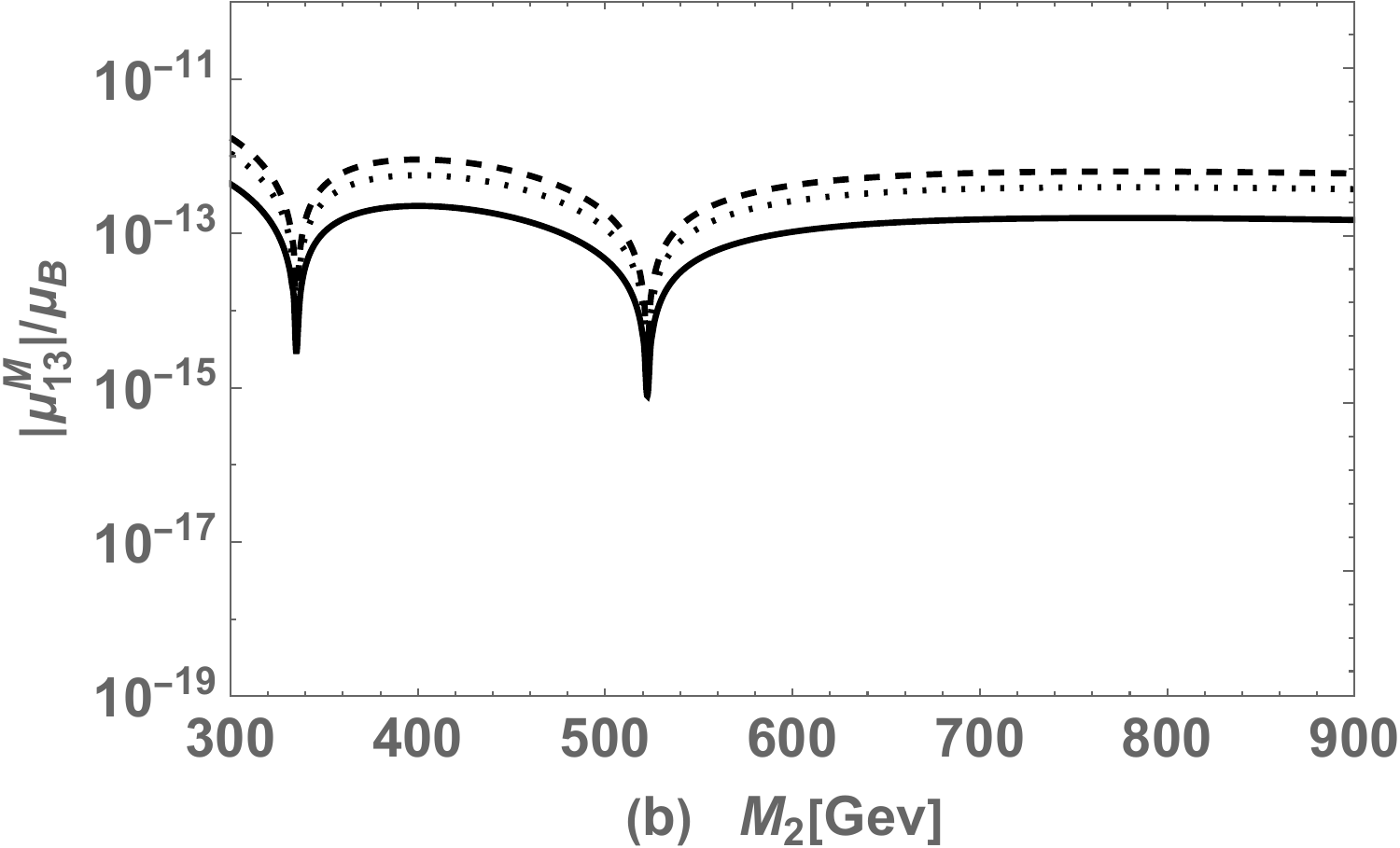}
	\end{minipage}%
	\begin{minipage}[c]{0.5\textwidth}
		\includegraphics[width=2.9in]{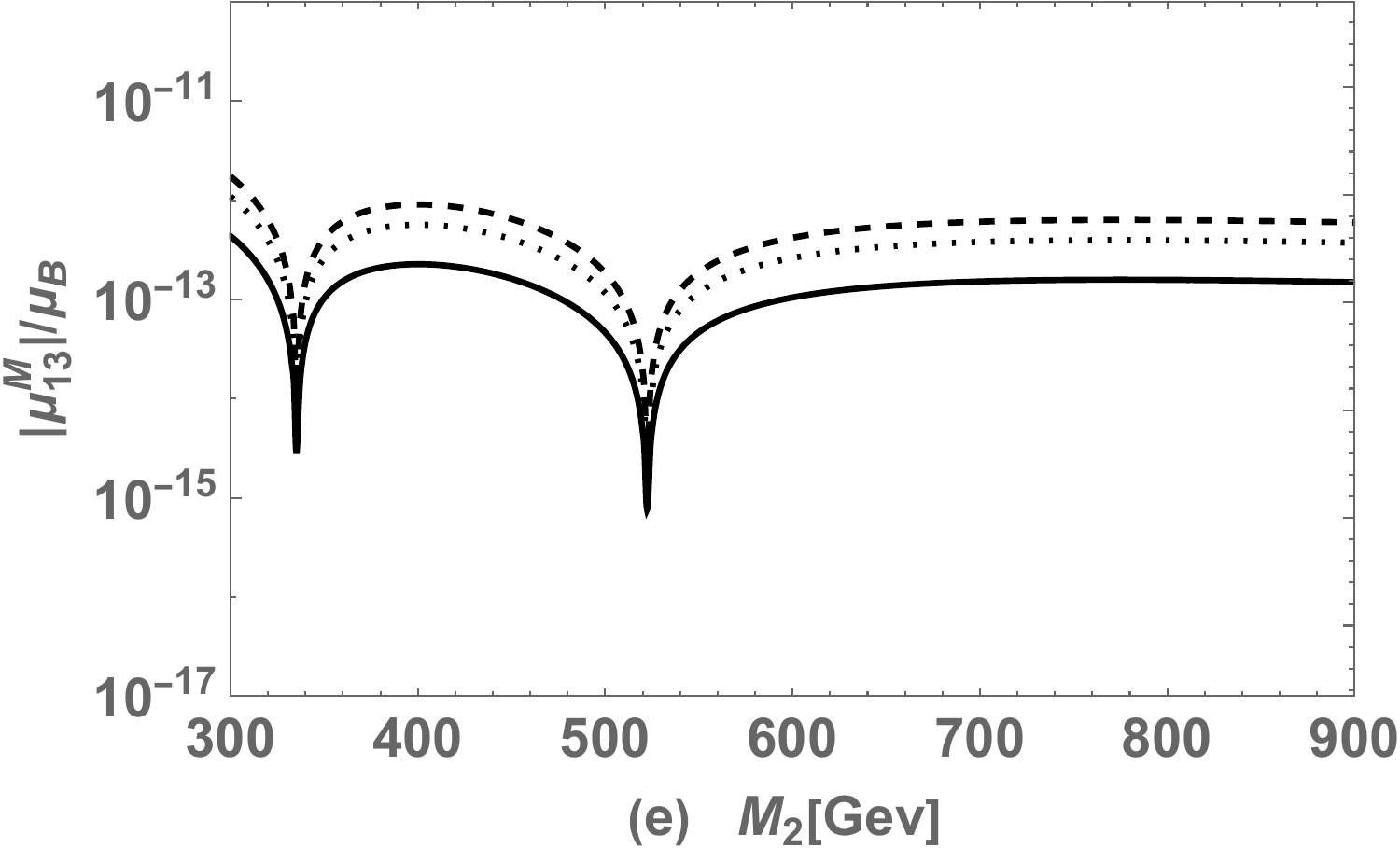}
	\end{minipage}
	\begin{minipage}[c]{0.5\textwidth}
		\includegraphics[width=2.9in]{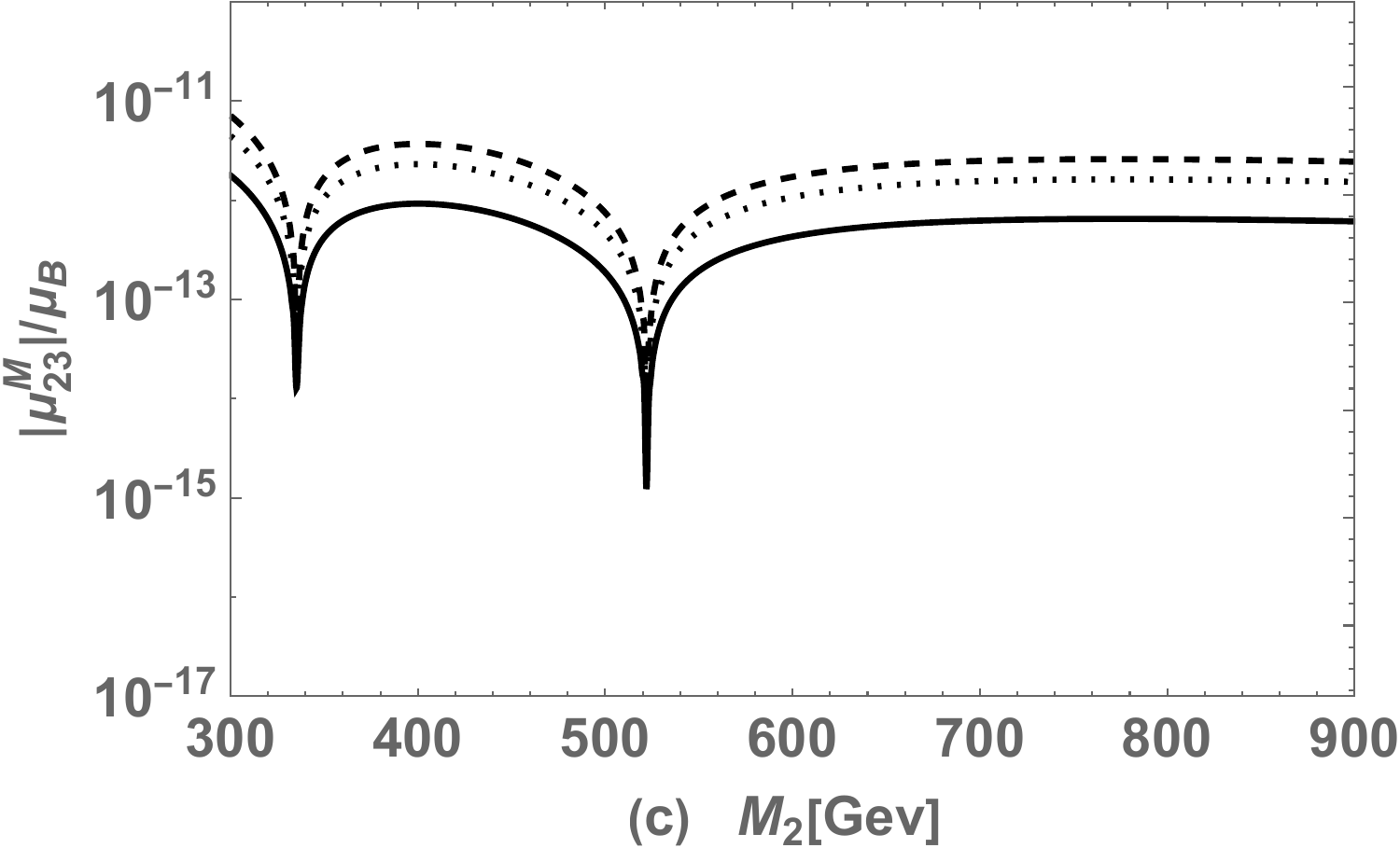}
	\end{minipage}%
	\begin{minipage}[c]{0.5\textwidth}
		\includegraphics[width=2.9in]{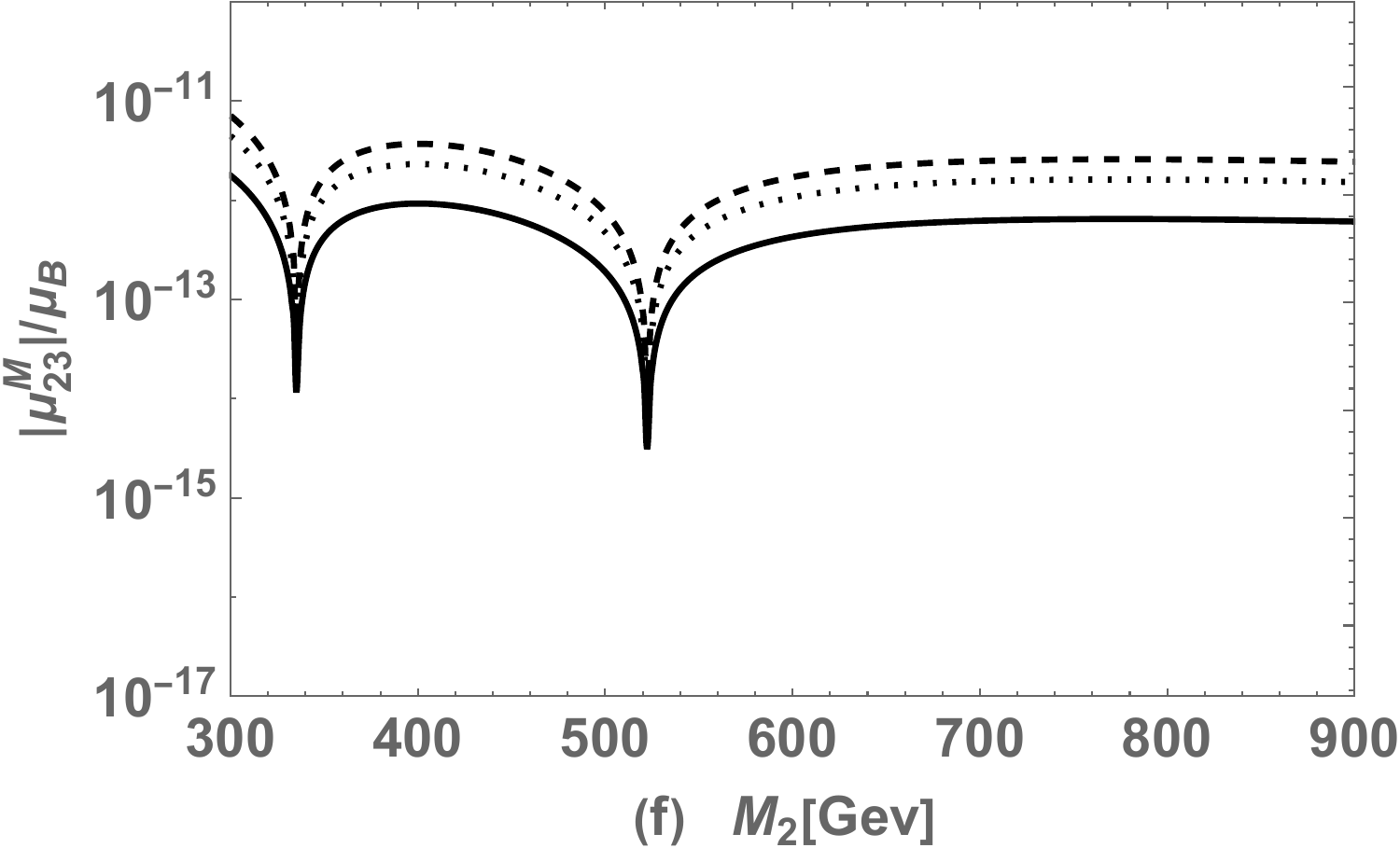}
	\end{minipage}

	\caption{The transition magnetic moment versus $M_2$ is plotted for NH (a, b, c) and IH (d, e, f) neutrino masses, where the solid, dotted, dashed lines represent $Y_{LL}=0.01$, $Y_{LL}=0.025$ and $Y_{LL}=0.04$ respectively. Here (a), (b), and (c) ((d), (e), and (f)) denote that $|\mu_{12}^M|/\mu_B$, $|\mu_{13}^M|/\mu_B$, and $|\mu_{23}^M|/\mu_B$, respectively, when the neutrino mass spectrum is NH (IH).}
	\label{M2SS}
\end{figure}
Without losing generality, we set $\tan \beta^{'}=0.8$ in later analysis of the transition magnetic moments of Majorana neutrinos.
In combination with the results of one-loop correction to the neutrino mass, we study the transition magnetic moments of Majorana neutrinos within the range allowed by the neutrino oscillation experiment.
In Fig.~\ref{M2SS}, we take $\mu=650 \;{\rm GeV}$, for NH (a, b, c) and IH (d, e, f) neutrino masses, then plot the transition magnetic moment versus $M_2$ for left-handed Majorana neutrinos, assuming that the neutrino mass spectrum with NH or IH. The solid, dotted, dashed lines represent $Y_{LL}=0.01$, $Y_{LL}=0.025$ and $Y_{LL}=0.04$ respectively. Fig.~\ref{M2SS} (a), (b), and (c) ((d), (e), and (f)) represent $|\mu_{12}^M|/\mu_B$, $|\mu_{13}^M|/\mu_B$, and $|\mu_{23}^M|/\mu_B$ results, respectively. We can find that the general trend on the left-handed Majorana neutrino transition magnetic moment decreases with increasing $M_2$. Here the soft breaking wino mass $M_2$ influences the wino-like chargino masses. As $M_2$ increases, the wino-like chargino mass also increases, leading to suppression of the wino-like chargino loop contribution. In addition, it is found that the general trend of the transition magnetic moment of Majorana neutrinos increases with $Y_{LL}$. Here $Y_{LL}$ affects $Y_{D}$ through Eq.~(\ref{MLR}) and Eq.~(\ref{meff1}). To obtain the tiny neutrino masses when $Y_{LL}$ increases the $Y_{D}$ also increases. Both $Y_{LL}$ and $Y_{D}$ can affect the interaction of neutrinos with fermions and scalars. This is because large $Y_{LL}$ and $Y_{D}$ enhance the couplings between neutrinos with fermions and scalars, which give significant contributions to the transition magnetic moment through the charged fermions and charged scalars loop diagrams.

However, in Fig.~\ref{M2SS} we can also see that the transition magnetic moments of Majorana neutrinos may drop sharply in certain parameter spaces. This is due to the fact that the Majorana neutrino coincides with its antiparticle, the Majorana neutrino transition magnetic moment $\mu_{ij}^M$ in Eq.~(\ref{Majorana-MDM}) contains the Dirac-neutrino-like term $\mu_{ij}^D$ and the Dirac-antineutrino-like term $-\mu_{ji}^D$. Considered the supersymmetric particle loop contributions in the TNMSSM, the Majorana neutrino transition magnetic moment may have resonant absorption in some parameter space, which originates from the interference between the Dirac-neutrino class term and the Dirac-antineutrino class term.

In Fig.~\ref{M2SS}  shows that there are some differences between the two different types of neutrino mass spectra, with the NH changing more smoothly than the IH. Thus in the future, with a more precise understanding of the transition magnetic moments of Majorana neutrinos, it may be possible to differentiate the hierarchy of the neutrino mass.

	\begin{figure}
		\setlength{\unitlength}{1mm}
		\centering
		\begin{minipage}[c]{0.5\textwidth}
			\includegraphics[width=2.9in]{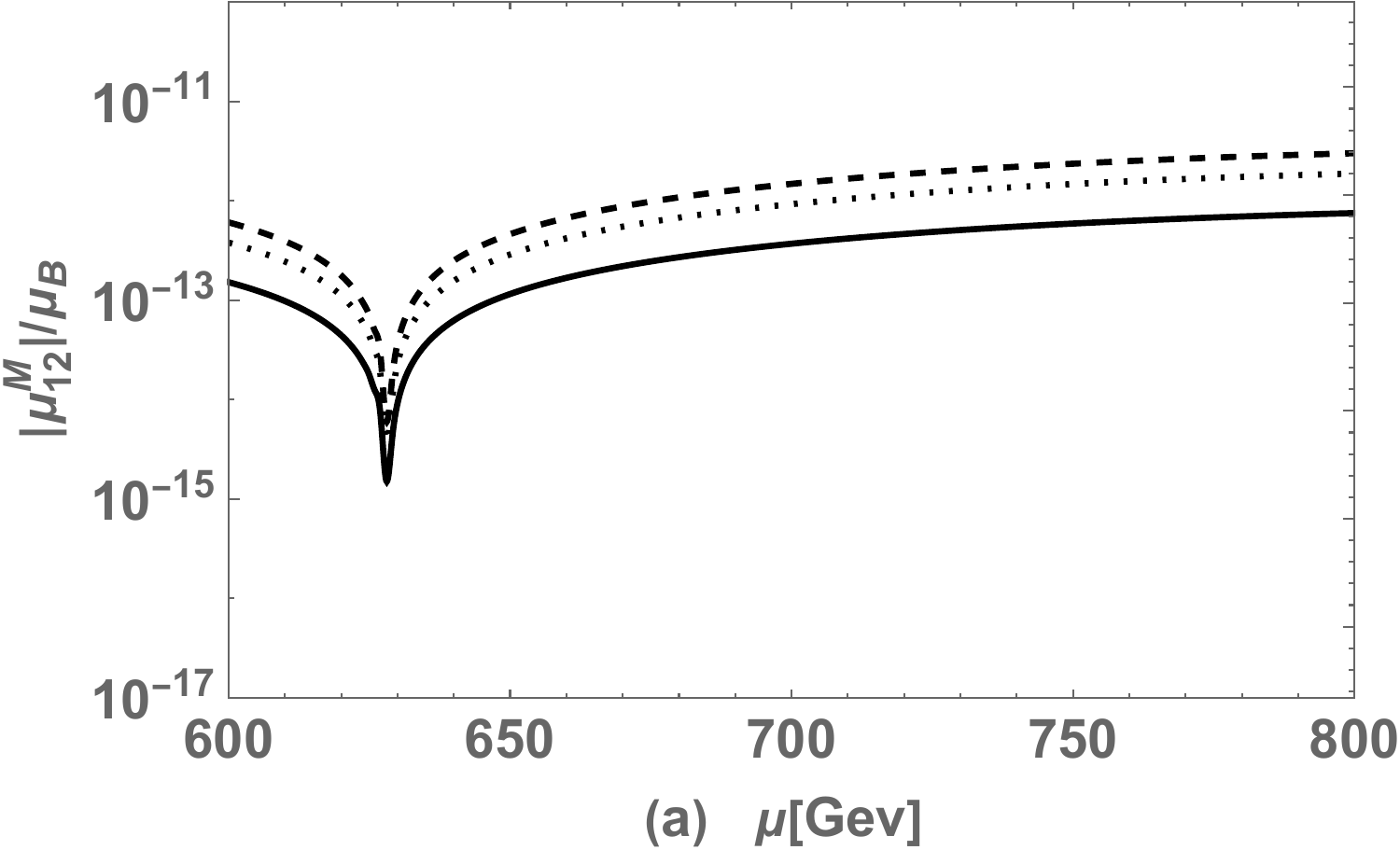}
		\end{minipage}%
		\begin{minipage}[c]{0.5\textwidth}
			\includegraphics[width=2.9in]{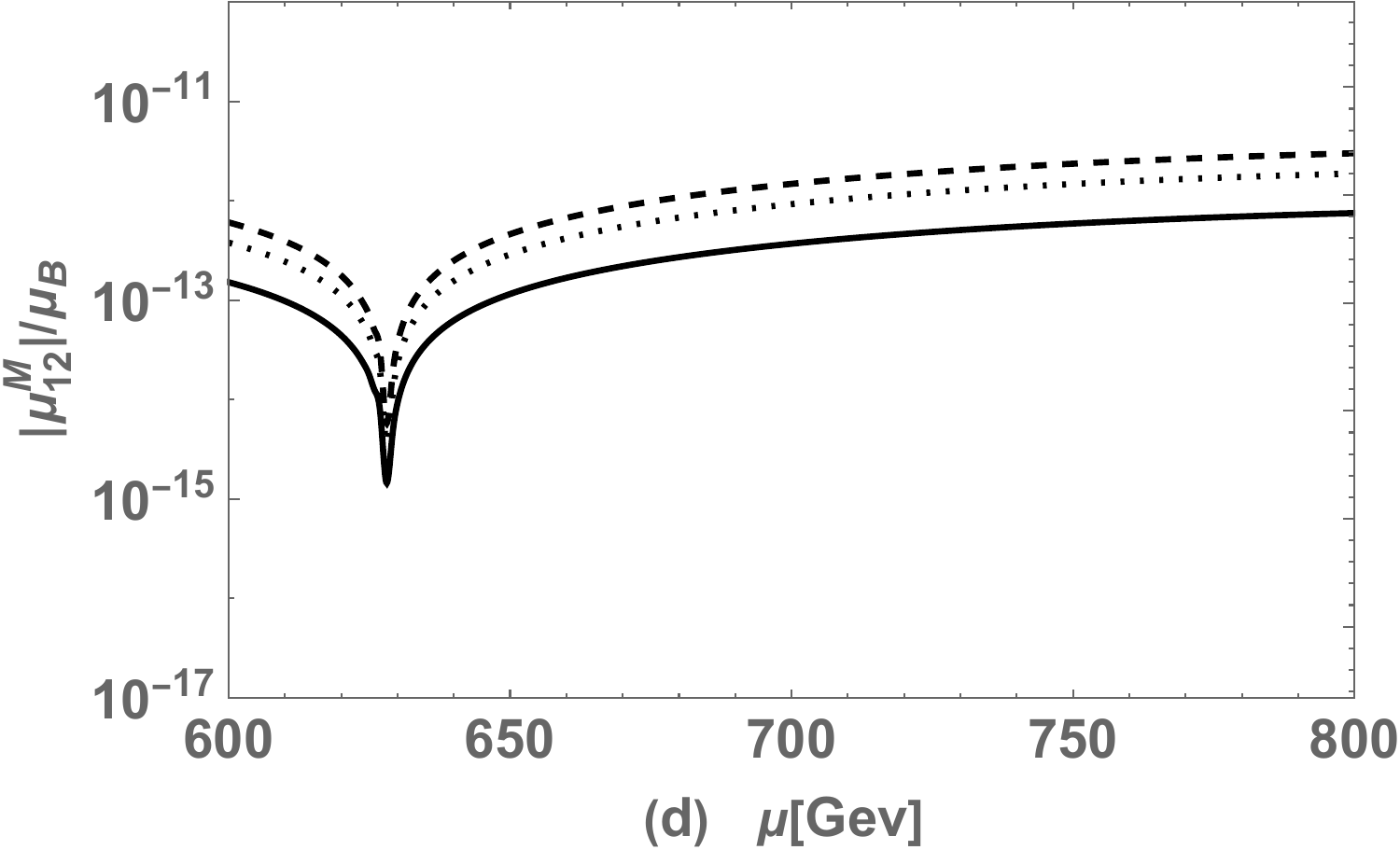}
		\end{minipage}
		\begin{minipage}[c]{0.5\textwidth}
			\includegraphics[width=2.9in]{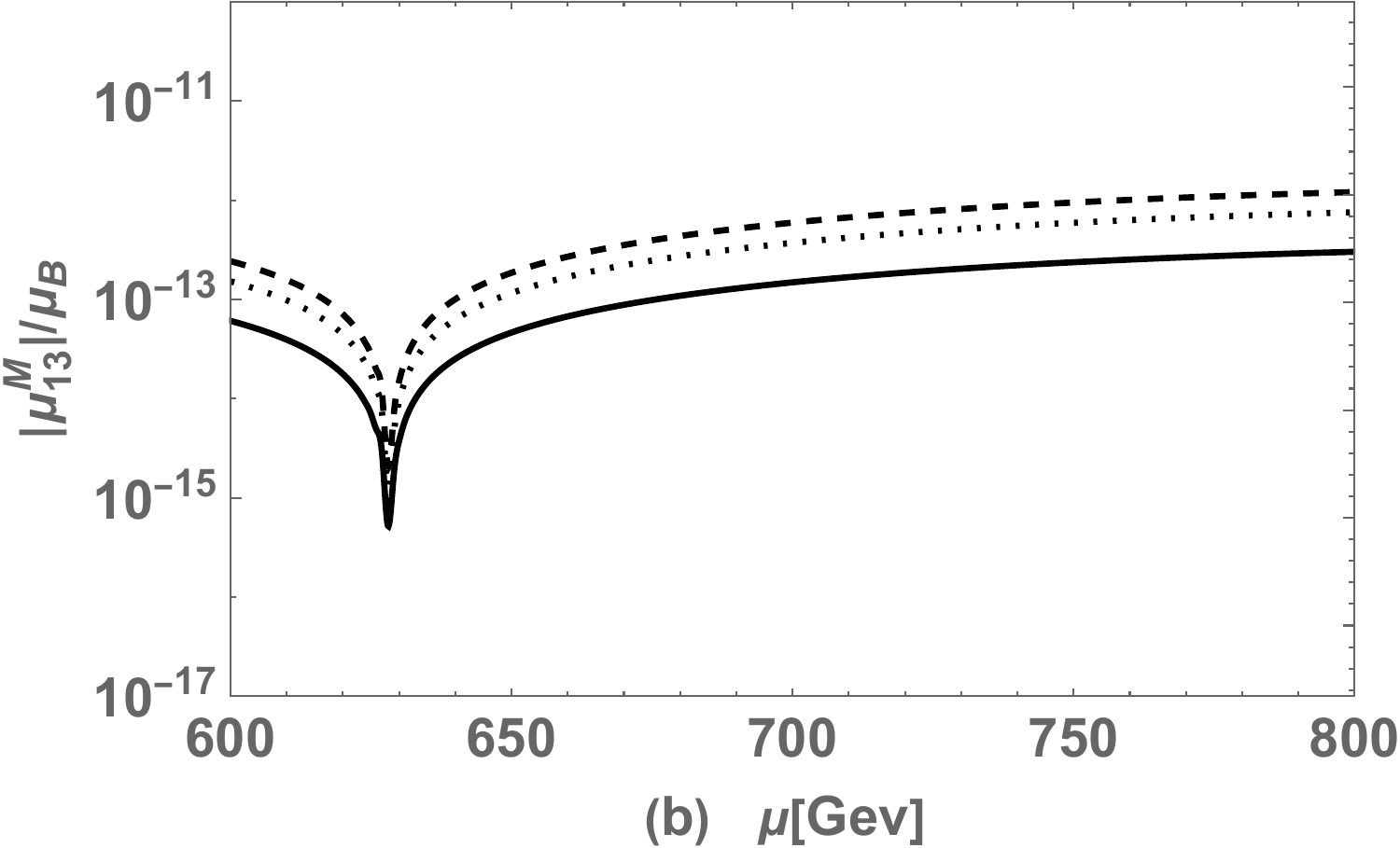}
		\end{minipage}%
		\begin{minipage}[c]{0.5\textwidth}
			\includegraphics[width=2.9in]{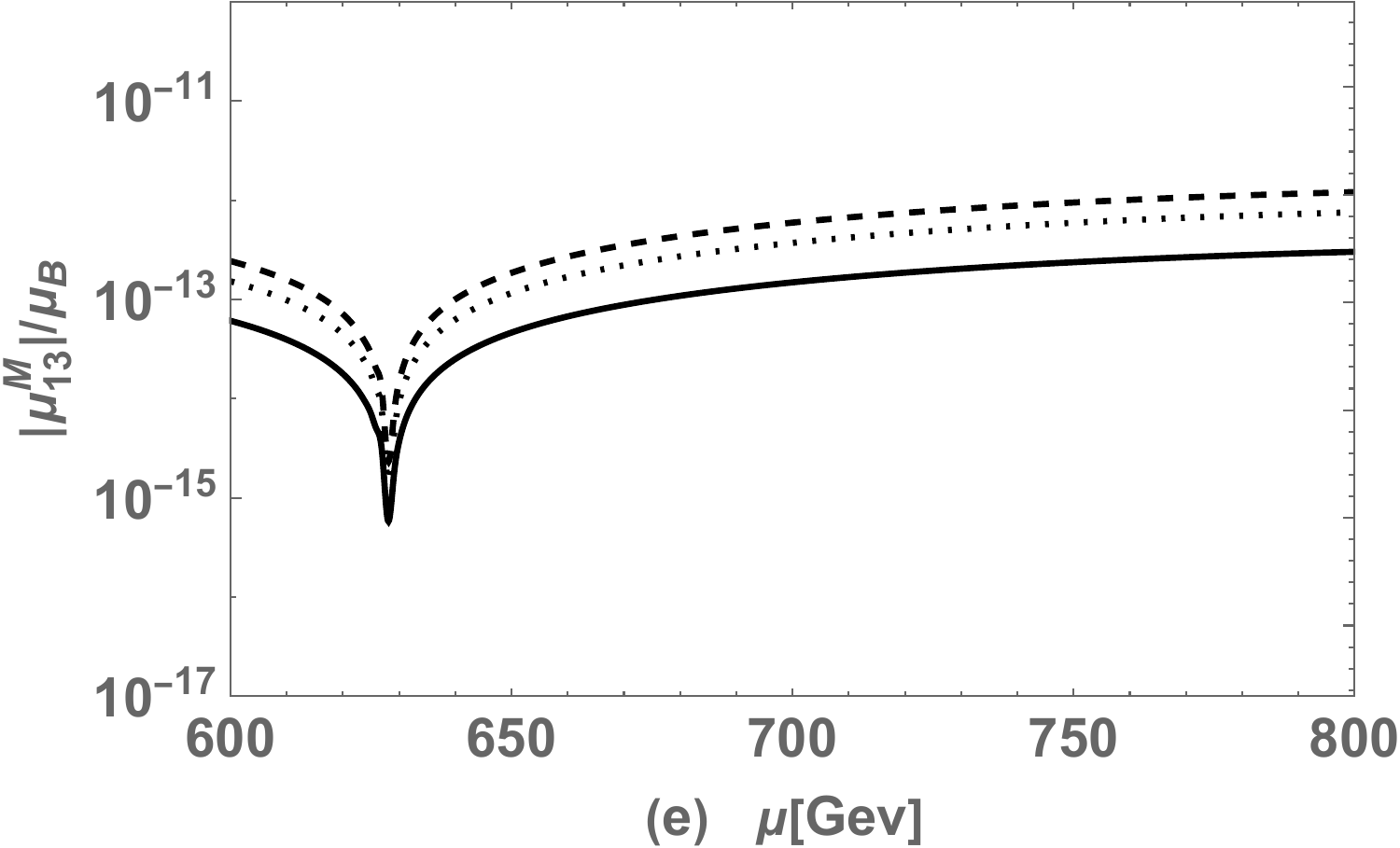}
		\end{minipage}
		\begin{minipage}[c]{0.5\textwidth}
			\includegraphics[width=2.9in]{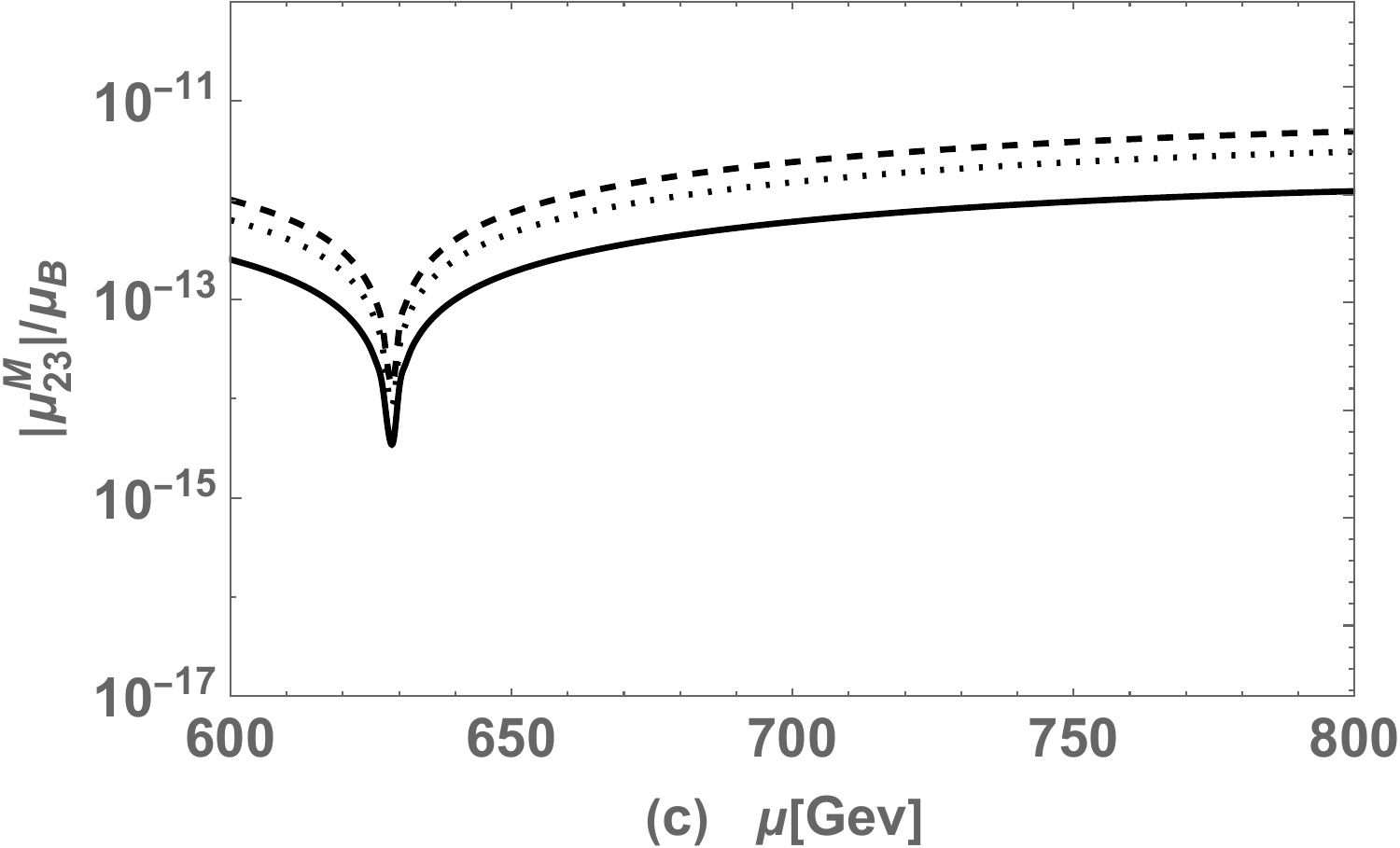}
		\end{minipage}%
		\begin{minipage}[c]{0.5\textwidth}
			\includegraphics[width=2.9in]{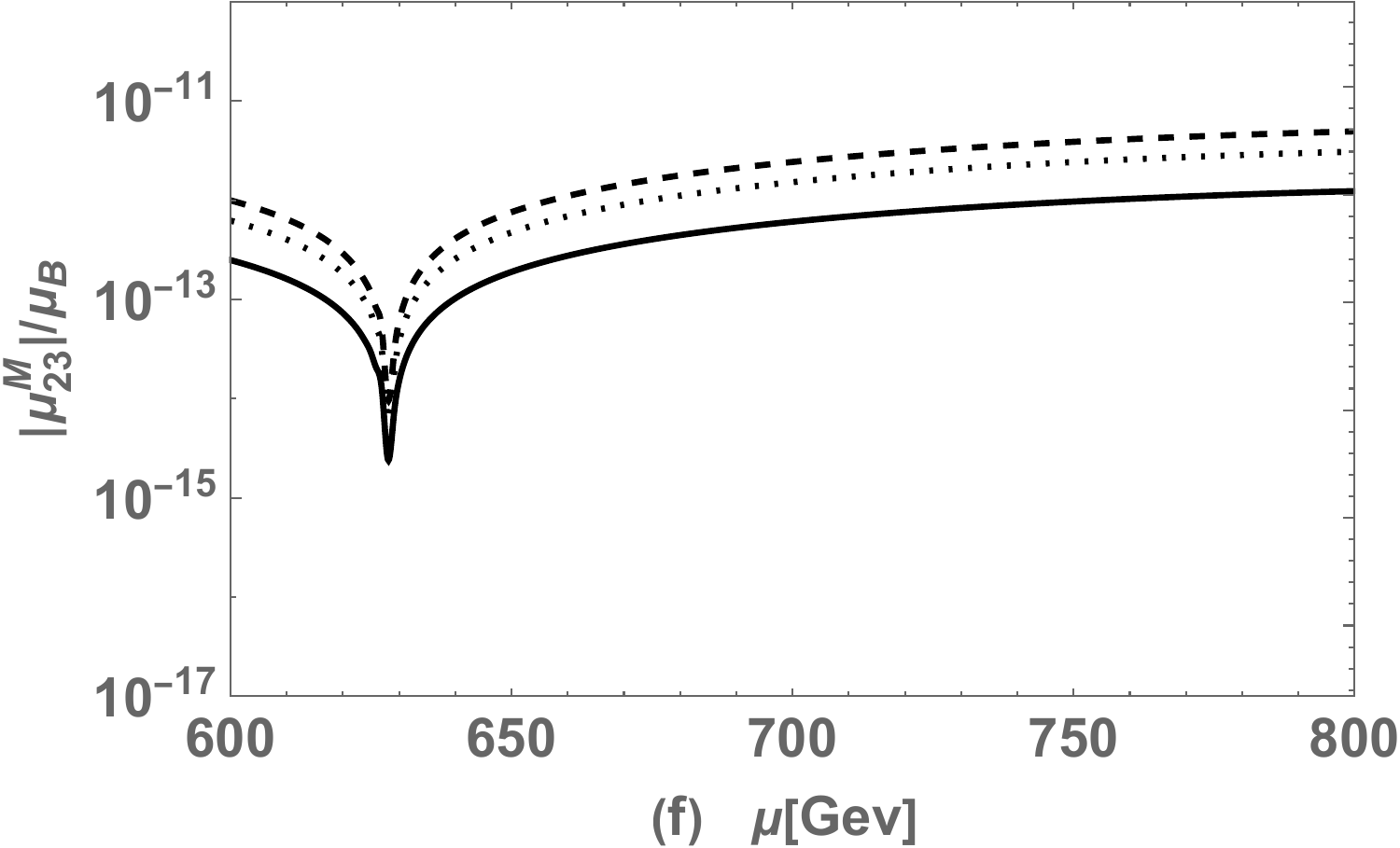}
		\end{minipage}

		\caption{The transition magnetic moment versus $\mu$ are plotted for NH (a, b, c) and IH (d, e, f) neutrino masses, where the solid, dotted, dashed lines represent $Y_{LL}=0.01$, $Y_{LL}=0.025$ and $Y_{LL}=0.04$ respectively. Here (a), (b), and (c) ((d), (e), and (f)) denote that $|\mu_{12}^M|/\mu_B$, $|\mu_{13}^M|/\mu_B$, and $|\mu_{23}^M|/\mu_B$, respectively, when the neutrino mass spectrum is NH (IH).}
		\label{YEEM2}
	\end{figure}
	
	In Fig.~\ref{YEEM2}, we take $M_2=500 \;{\rm GeV}$, for NH (a, b, c) and IH (d, e, f) neutrino masses. Then the transition magnetic moment of Majorana neutrinos varies with $\mu$ assuming neutrino mass spectrum with NH or IH, as $|\mu_{12}^M|/\mu_B$ (a,d),  $|\mu_{13}^M|/\mu_B$ (b,e) and $|\mu_{23}^M|/\mu_B$ (c,f), respectively. Where the solid, dotted, dashed lines denote the results of $Y_{LL}=0.01$, $Y_{LL}=0.025$ and $Y_{LL}=0.04$ respectively. It can be found that within the range given, the general trend of the transition magnetic moment of Majorana neutrinos increases with the increase of $\mu$ and with the increase of $Y_{LL}$. Here, when we fix $\lambda$ and $M_{R}$, through Eq.~(\ref{uuuu}), $v_S$ will also increase as $\mu$ increases, which affects the mass of the charged Higgs, the mass of the slepton, the wino-like chargino masses and the interaction of neutrinos with fermions and scalars $Y_{R}$. Therefore, the change of $\mu$ will have a significant effect on the transition magnetic moment of Majorana neutrinos. However, $Y_{R}$ should decrease as $\mu$ increases and the wino-like chargino masses should increase which would suppress the contribution of the loop diagram, but we did not see the expected result. For the same reason as $M_2$, the resonance absorption phenomenon also occurs with the change of $\mu$, and in a given range, it just leaves the resonance absorption point, so we can see that the transition magnetic moment also increases with the increase of $\mu$.

\begin{figure}
	\setlength{\unitlength}{1mm}
	\centering
	\includegraphics[width=3in]{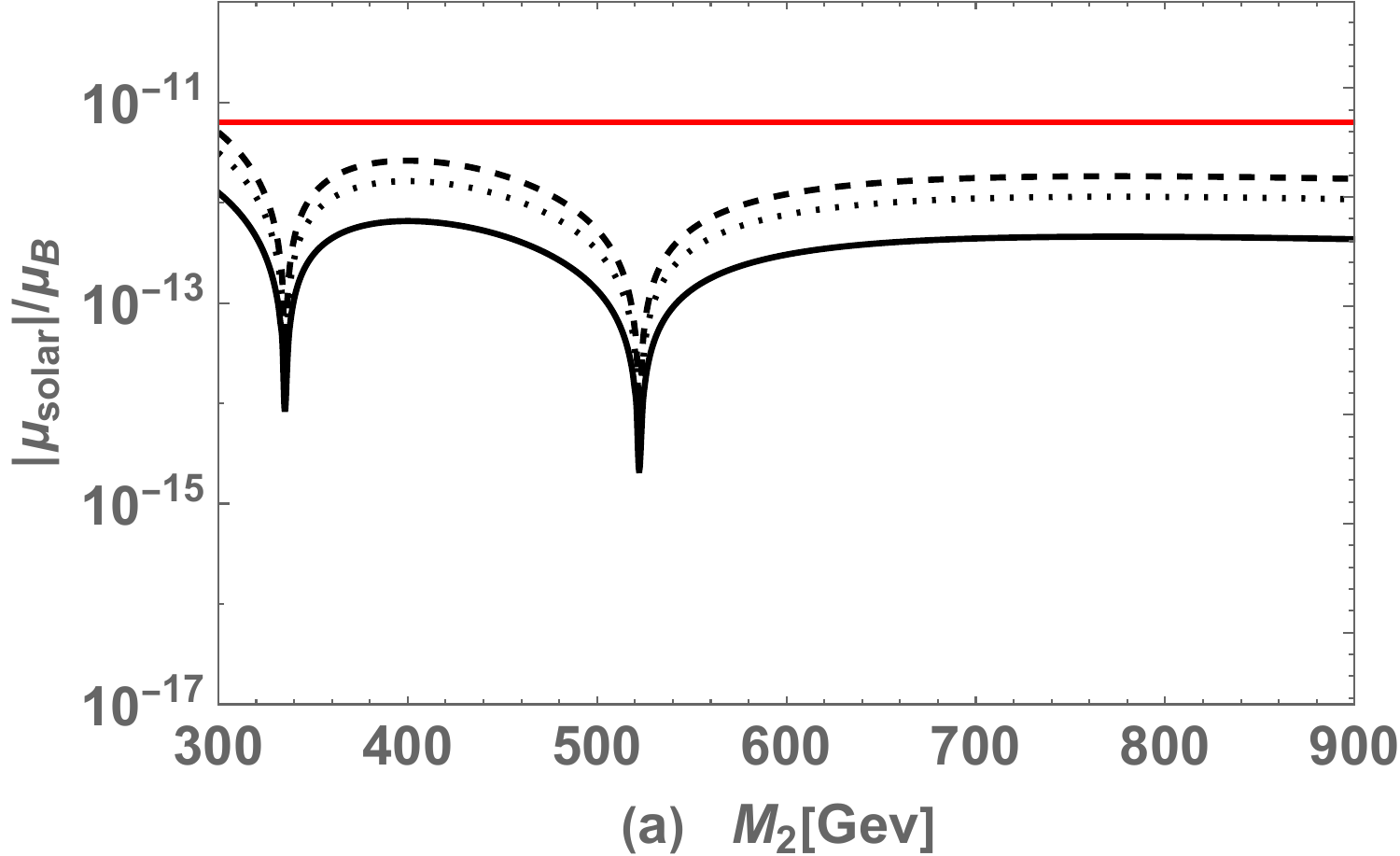}
	\vspace{0.5cm}
	\includegraphics[width=3in]{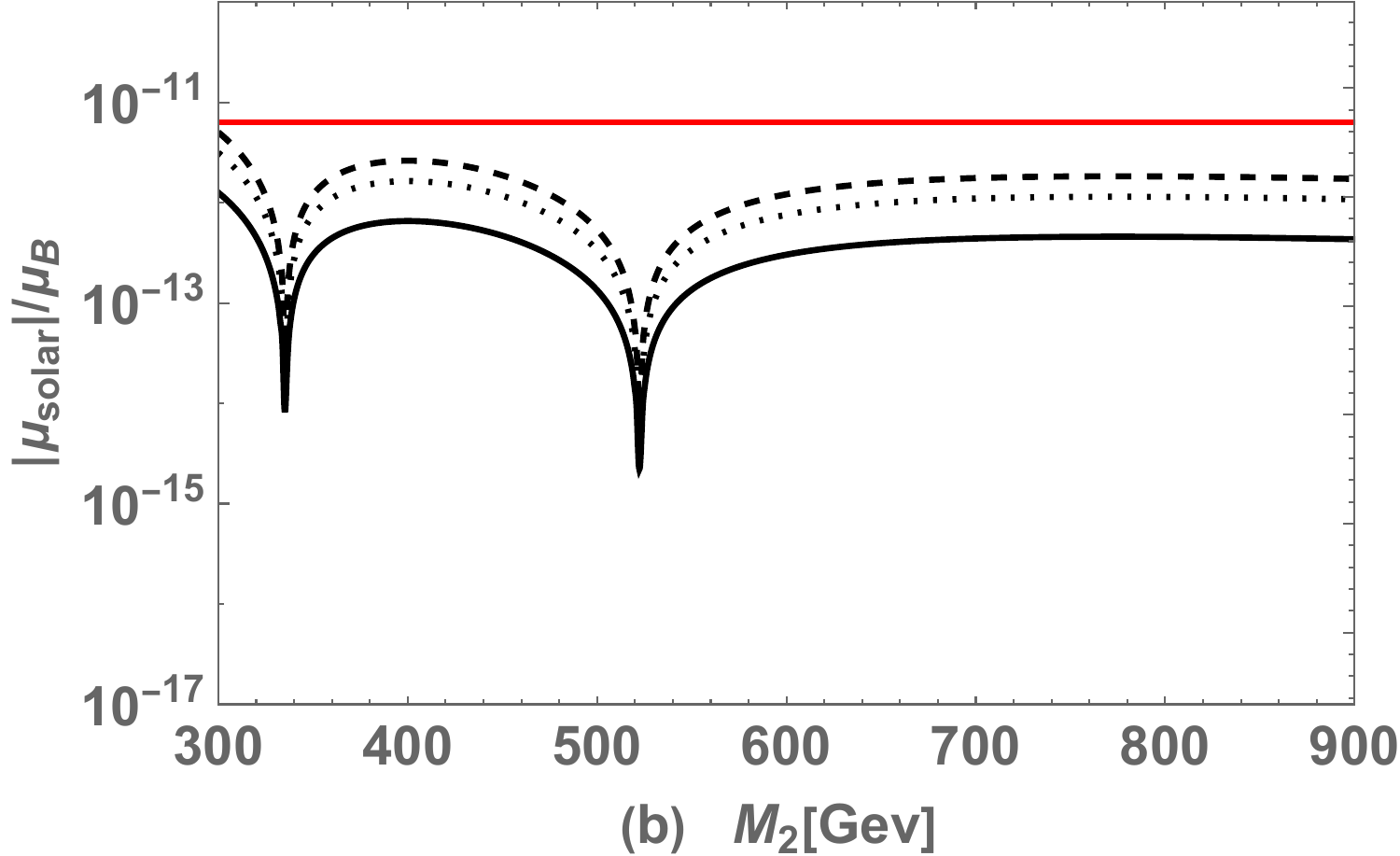}
	\vspace{0cm}
	\caption{$|\mu_{\nu {\rm solar}}|/\mu_B$ versus $M_{2}$ are plotted for NH (a) and IH (b) neutrino mass, where the solid, dashed, dotted lines denote the results of $Y_{LL}=0.01$, $Y_{LL}=0.025$ and $Y_{LL}=0.04$, respectively. The red solid line denotes the constraints from XENONnT experiment.}
	\label{SOA}
\end{figure}
Recent results from XENONnT have pushed the laboratory limit down to $6.3\times10^{-12}\mu_B$ at the $90\%$ C.L \cite{XTN7}. The upper limit on solar neutrinos with an enhanced transition magnetic moment is $6.3\times10^{-12}\mu_B$. For Majorana neutrinos, the relation takes the form \cite{zy2}
\begin{eqnarray}
	\mu^2_{\nu{\rm solar}} = |\mu_{12}|^2 c^2_{13} + |\mu_{13}|^2(c^2_{13} 
	\cos^2 \Tilde{\theta} + s^2_{13}) + 
	|\mu_{23}|^2(c^2_{13} \sin^2 \Tilde{\theta} + s^2_{13}).
	\text{}
	\label{eq:muMaj}
\end{eqnarray}
\begin{figure}
	\setlength{\unitlength}{1mm}
	\centering
	\includegraphics[width=3in]{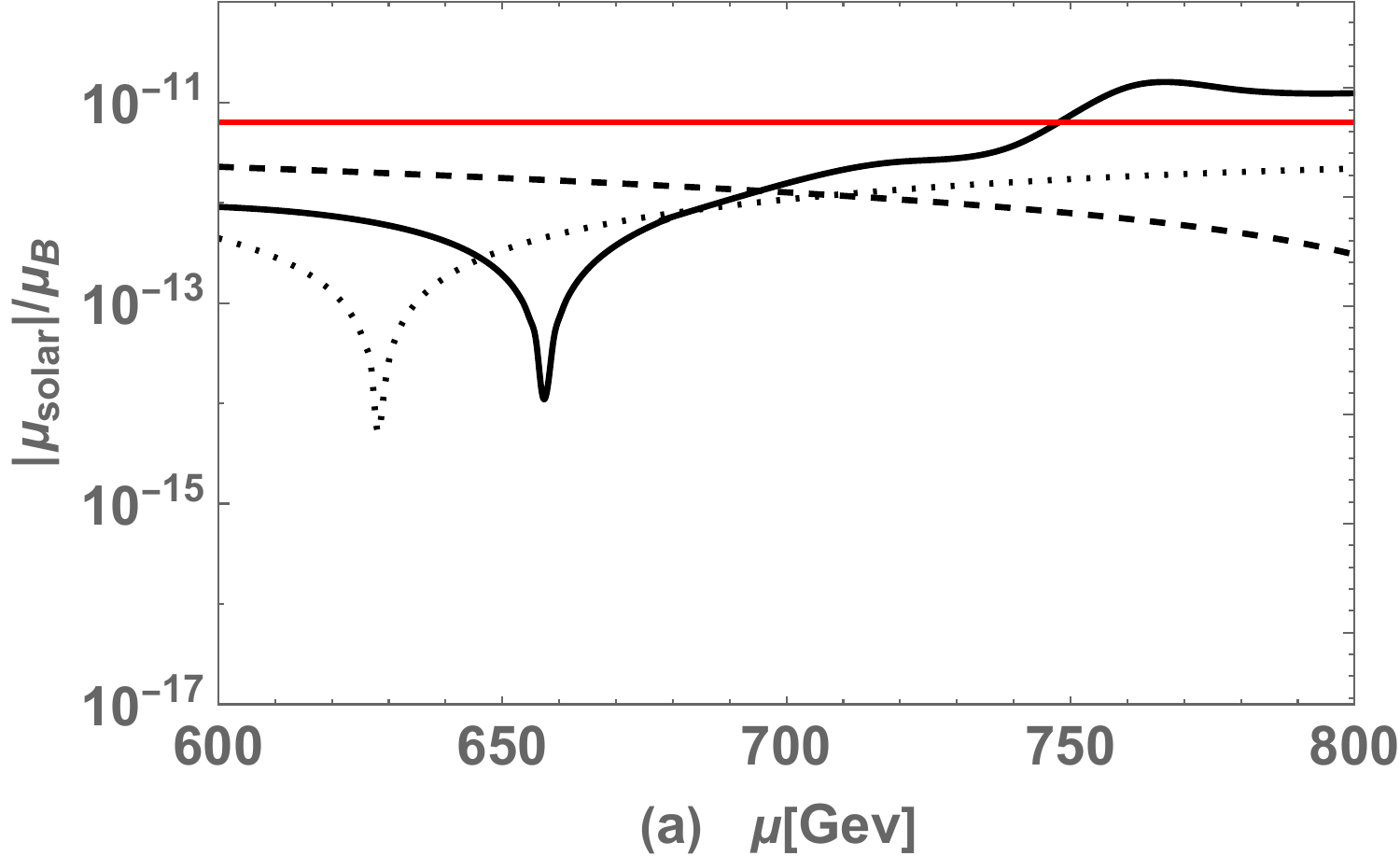}
	\vspace{0.5cm}
	\includegraphics[width=3in]{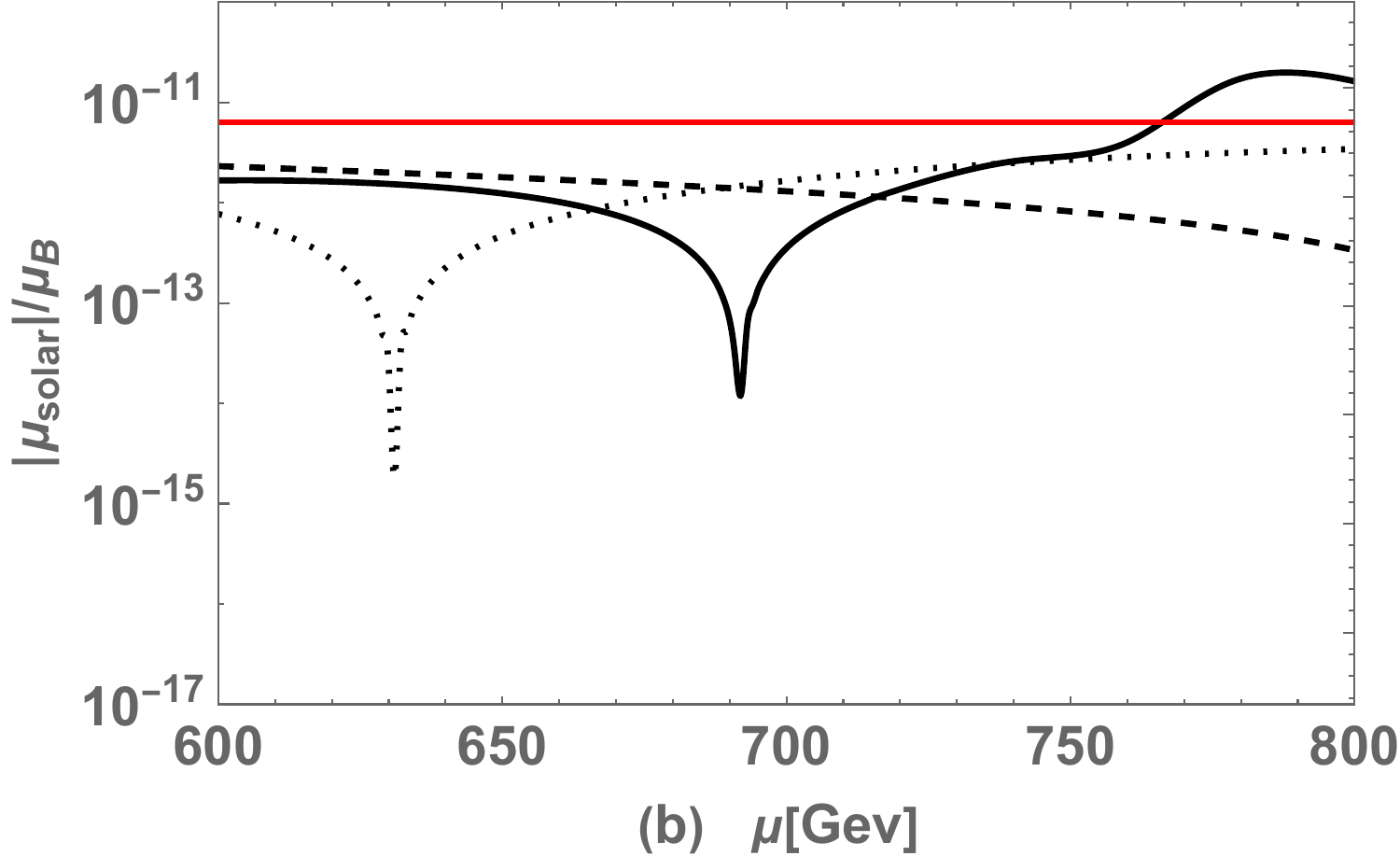}
	\vspace{0cm}
	\caption{$|\mu_{\nu {\rm solar}}|/\mu_B$ versus $\mu$ are plotted for NH (a) and IH (b) neutrino mass, where the solid, dashed, dotted lines denote the results of $M_{2}=300 \;{\rm GeV}$, $M_{2}=500 \;{\rm GeV}$ and $M_{2}=700 \;{\rm GeV}$, respectively. The red solid line denotes the constraints from XENONnT experiment.}
	\label{SOAY}
\end{figure}
Here $\mu_{\nu{\rm solar}}$ represents the solar neutrinos transition magnetic moment. We take  $\Tilde{\theta} \simeq \theta_{12}$ approximately. In Fig.~\ref{SOA}, $\mu=650 \;{\rm GeV}$ in NH (a) and IH (b), then plot the transition magnetic moment of the solar neutrinos versus $M_{2}$.  The solid, dashed and dotted lines are the results for $Y_{LL}=0.01$, $Y_{LL}=0.025$ and $Y_{LL}=0.04$ respectively. The red solid line denotes the constraints from XENONnT experiment. It is found that for smaller values of $M_{2}$ and larger values of $Y_{LL}$, relatively large solar neutrino transition magnetic moments can be obtained. Since the $\mu_{\nu{\rm solar}}$ is a combination of $|\mu_{12}^M|/\mu_B$, $|\mu_{13}^M|/\mu_B$, $|\mu_{23}^M|/\mu_B$, there may be a resonance absorption of the solar neutrino transition magnetic moments, which also originates from the interference between the Dirac-neutrino class term and the Dirac-antineutrino class term. In addition, In Fig.~\ref{SOAY}, we take $Y_{LL}=0.04$ in NH (a) and IH (b), then plot the transition magnetic moment of the solar neutrinos versus $\mu$.  The solid, dashed and dotted lines are the results for $M_{2}=300 \;{\rm GeV}$, $M_{2}=500 \;{\rm GeV}$ and $M_{2}=700 \;{\rm GeV}$ respectively. $\mu$ can influence the couplings between neutrinos with fermions and scalars by influencing $v_S$, and can also influence the masses of particles including chargino, charged Higgs, slepton, and others. Therefore, the change of $\mu$ will have a significant effect on the solar neutrino transition magnetic moment. We can see that with the change of $M_{2}$, the solar neutrino transition magnetic moments also change accordingly. Compared with Fig.~\ref{YEEM2}, the resonance absorption point of $\mu$ moves with the change of $M_{2}$. And, when $M_{2}=320 \;{\rm GeV}$ and $\mu$ is greater than $750 \;{\rm GeV}$, the transition magnetic moment of the solar neutrinos will exceed the limit given by the XENONnT experiment. In addition, it can be found that when $M_{2}=700 \;{\rm GeV}$, it can be found that  the solar neutrino transition magnetic moments decreases correspondingly with the increase of $\mu$. This gives rise to the expected phenomenon that large particle masses will depress the contribution of the loop diagrams. Moreover, we can find that the resonance absorption points are not the same for different neutrino mass spectrum, and perhaps in the future, we can indirectly give a evidence to explain the neutrino mass spectrum by in-depth study of the transition magnetic moment. From Fig.~\ref{SOA} and Fig.~\ref{SOAY}, we can see that the transition magnetic moments of solar neutrinos will exceed the limit given by XENONnT experiment in some parameter space. This will further limit our parameter space, and improve our understanding of the neutrino transition magnetic moment, with better reference value for future research. With the further development of experimental physics, more stringent limits on the neutrinos transition magnetic moment will be proposed, further limiting the parameter space.  Thus improving the understanding of neutrino magnetic moments may indirectly improve the understanding of the $0\nu2\beta$ decay experiments and neutrino mass generation mechanism, the hierarchy of neutrino mass spectra, as well as the new physics \cite{zy3}.
	\section{Conclusions}
	\label{sec5}
	In the paper, we take the next-to-minimal Supersymmetric Standard Model with triplets (TNMSSM) of new physics models to study the transition magnetic moment of Majorana neutrinos. The model gives three tiny Majorana neutrino masses via the type I+II seesaw mechanism. The neutrino mass matrix of the tree-level is given by the type I+II seesaw mechanism. Further, we use the on-shell scheme to consider the effect of the one-loop corrections on the tree-level neutrino mass matrix. Then the mass matrix of neutrinos is given by combining the results of tree-level and one-loop. Due to the nature of type I+II seesaw mechanism, the smallness of $M_{\nu}^{eff} \approx (M_L+\delta(m_L)) -
	(M_D+\delta(m_D))\cdot{(M_R+\delta({m_R})))}^{-1}\cdot{(M_D+\delta(m_D))}$. is attributed to a
	significant but incomplete cancellation between $M_L+\delta(m_L)$ and $(M_D+\delta(m_D))\cdot{(M_R+\delta({m_R})))}^{-1}\cdot{(M_D+\delta(m_D))}$ terms. Applying the effective Lagrangian method and on-shell scheme, we investigate the transition magnetic moment of the Majorana neutrino in the model. Under the constraints of the current experimental data on neutrino physics and some assumptions of parameter space, we consider the mass spectrum of neutrinos with two possibilities, NH neutrino masses and IH neutrino masses.

When the neutrino mass includes one-loop correction, numerical results show that the effects of the one-loop correction increase as the $Y_{LL}$ increases, because they can affect the interaction of neutrinos with fermions and scalar bosons neutrinos through type I+II seesaw mechanism, which in turn affects the neutrino mass. Then we found that the sneutrino-neutralino and the slepton-chargino diagram made a significant contribution, perhaps indicating a deeper connection between neutrino and neutralino and chargino. The contribution of the one-loop correction decreases as the mass of neutralinos and charginos increases. We adjust the parameter space according to the neutrino experiment to ensure that the neutrino mass given by the tree-level and one-loop correction is within the range allowed by the neutrino experiment. In addition,
the numerical results show that when the supersymmetric particles are light and the couplings between neutrinos with charged fermions and charged scalars are large, the transition magnetic moment of Majorana neutrinos in the TNMSSM can be enhanced to $\mathcal{O}(10^{-11}\mu_B)$. The mass of the tiny neutrino is in the range given by the neutrino oscillation experiment, and at the same time, a large transition magnetic moment is obtained.
The masses of particles in the charged fermions and charged scalar loop diagrams have significant effects on the transition magnetic moment of the Majorana neutrinos. Resonant absorption may occur in some parameter spaces due to interference between Dirac-like neutrino terms and Dirac-like antineutrino terms.		 
We compare the calculated solar neutrino transition magnetic moments with the latest XENONnT experimental results and find that the results will exceed the XENONnT experimental limits in some parameter spaces. This will further limit the current parameter space, provide rich phenomenology, and may have some reference value for future research. In addition, the in-depth study of neutrino transition magnetic moments, may indirectly give evidence to explain the mass order of neutrinos.  This model is simple but phenomenologically offers rich content for a solution to neutrino properties. A deeper understanding of the Majorana neutrino transition magnetic moment in the future may indirectly lead to a further understanding of neutrino properties and the neutrino
mass generation mechanisms as well as new physics.

	\section*{Acknowledgements}
	The work has been supported by Natural Science Foundation of Guangxi Autonomous Region with Grant No. 2022GXNSFDA035068, the National Natural Science Foundation of China (NNSFC) with Grants No. 12075074, No. 12235008, No. 11535002, No. 11705045, Hebei Natural Science Foundation for Distinguished Young Scholars with Grant No. A2022201017,  and the youth top-notch talent support program of the Hebei Province.

\appendix

\section{The one-loop corrections to the neutrino masses in the TNMSSM\label{oloop}}	

In this section, we compute the one-loop corrections to the neutrino masses in the TNMSSM. The general form of the self-energy for ${\nu_ i}^0-{\nu_ j}^0$ can be written
as \cite{onel2}	
\begin{eqnarray}
	&&\Sigma(k)_{ij}=c_{ij}m_{j}\omega_-+d_{ij}m_{i}\omega_++e_{ij}
	/\!\!\! k\omega_-+f_{ij}/\!\!\! k \omega_+.
	\label{onel1}
\end{eqnarray}

When the external leg of the self-energy is neutrino,  $k^2\ll m_0^2$
with $m_0$ being the mass of the heaviest internal particle and $c_{ij}$, $d_{ij}$, $e_{ij}$ and $f_{ij}$ can be written as an expansion of $k^2$\cite{yao}:
\begin{eqnarray}
	&&c_{ij}=c_{ij}^0+k^2c_{ij}^1, \!\! \nonumber\\
	&&d_{ij}=d_{ij}^0+k^2d_{ij}^1, \!\! \nonumber\\
	&&e_{ij}=e_{ij}^0+k^2e_{ij}^1, \!\! \nonumber\\
	&&f_{ij}=f_{ij}^0+k^2f_{ij}^1. \!\!
	\label{series}
\end{eqnarray}

$\Sigma_{ij}$'s are renormalized by adding counter-terms and the renormalized
$\Sigma_{ij}^\text{{Ren}}$ are written as:
\begin{equation}
	\Sigma_{ij}^\text{{Ren}}(k)=\Sigma_{ij}(k)+\Big(c_{ij}^*m_{j}\omega_-+d_{ij}^*m_i
	\omega_++e_{ij}^*/\!\!\!k\omega_-+f_{ij}^*/\!\!\!k\omega_+\Big),
	\label{counter}
\end{equation}

where the quantities with * are the counter parts.
In the on-shell renormalization scheme, it is determined by the mass-shell
conditions
\begin{eqnarray}
	\Sigma^\text{{Ren}}_{ij}(k)u_i(k)|_{k^2=m_i^2}=0,\\\nonumber
	\bar{u}_{j}(k)\Sigma^\text{{Ren}}_{ij}(k)|_{k^2=m_j^2}=0,
	\label{shell}
\end{eqnarray}
the solution can be written as
\begin{eqnarray}
	&&c_{ij}^*=-c_{ij}^0+m_i^2d_{ij}^1+m_i^2e_{ij}^1+m_im_jf_{ij}^1,\nonumber \\
	&&d_{ij}^*=-d_{ij}^0+m_j^2c_{ij}^1+m_j^2e_{ij}^1+m_im_jf_{ij}^1,\nonumber \\
	&&e_{ij}^*=-e_{ij}^0-m_j^2c_{ij}^1-m_i^2d_{ij}^1-(m_i^2+m_j^2)e_{ij}^1
	-m_im_jf_{ij}^1,\nonumber \\
	&&f_{ij}^*=-f_{ij}^0-m_im_jc_{ij}^1-m_im_jd_{ij}^1-m_im_je_{ij}^1
	-(m_i^2+m_j^2)f_{ij}^1.
	\label{solu1}
\end{eqnarray}
From Eq.~(\ref{counter}) and Eq.~(\ref{solu1}), the renormalized self-energy can be written as:
\begin{eqnarray}
	&&\Sigma^\text{{Ren}}_{ij}(k)=\Big(m_i^2d_{ij}^1+m_i^2e_{ij}^1+m_im_jf_{ij}^1
	+c_{ij}^1k^2\Big)m_j\omega_- \nonumber \\
	&&\hspace{1.5cm}+\Big(m_j^2c_{ij}^1+m_j^2e_{ij}^1+m_im_jf_{ij}^1
	+d_{ij}^1k^2\Big)m_i\omega_+ \nonumber \\
	&&\hspace{1.5cm}+\Big(-m_j^2c_{ij}^1-m_i^2d_{ij}^1-(m_i^2+m_j^2)e_{ij}^1
	-m_im_jf_{ij}^1+e_{ij}^1k^2\Big)/\!\!\!k\omega_- \nonumber \\
	&&\hspace{1.5cm}+\Big(-m_im_jc_{ij}^1-m_im_jd_{ij}^1-m_im_je_{ij}^1
	-(m_i^2+m_j^2)e_{ij}^1+f_{ij}^1k^2\Big)/\!\!\!k\omega_+\nonumber \\
	&&\hspace{1.5cm}=(/\!\!\!k-m_j)\hat{\Sigma}_{ij}(k)(/\!\!\!k-m_i).
	\label{solu2}
\end{eqnarray}

In the final step, $\Sigma_{ij}^\text{{Ren}}(k)$ was written to make its on-shell behavior more obvious as

\begin{eqnarray}
	&&\hat{\Sigma}_{ij}(k)=c_{ij}^1m_j\omega_{+}+d_{ij}^1m_i\omega_-
	+e_{ij}^1(m_i\omega_-+m_j\omega_++/\!\!\!k\omega_+)\nonumber \\
	&&\hspace{1.5cm}+f_{ij}^1(m_i\omega_++m_j\omega_-+/\!\!\!k\omega_-).
	\label{hatsig}
\end{eqnarray}
For convenience, some new symbols are introduced:
\begin{eqnarray}
	&&\delta Z_{ij}^{L}=-m_j^2c_{ij}^1-m_i^2d_{ij}^1-(m_i^2+m_j^2)e_{ij}^1
	-m_im_jf_{ij}^1+e_{ij}^1k^2,\nonumber \\
	&&\delta Z_{ij}^R=-m_im_jc_{ij}^1-m_im_jd_{ij}^1-m_im_je_{ij}^1
	-(m_i^2+m_j^2)f_{ij}^1+f_{ij}^1k^2,\nonumber \\
	&&\delta m_{ij}^{L}=\Big(m_i^2d_{ij}^1+m_i^2e_{ij}^1+m_im_jf_{ij}^1
	+c_{ij}^1k^2\Big)m_j,\nonumber \\
	&&\delta m_{ij}^R=\Big(m_j^2c_{ij}^1+m_j^2e_{ij}^1+m_im_jf_{ij}^1
	+d_{ij}^1k^2\Big)m_i.
	\label{symbel}
\end{eqnarray}
Up to one-loop order, the two-point Green  function is
\begin{eqnarray}
	&&\Gamma_{ij}(k)=\Big(/\!\!\!k-m_i^\text{{tree}}\Big)\delta_{ij}+\Sigma_{ij}^\text{{Ren}}(k)
	\nonumber \\
	&&\hspace{1.0cm}=\Big(/\!\!\!k-m_i^\text{{tree}}\Big)\delta_{ij}+\delta Z_{ij}^{L}
	/\!\!\!k\omega_-+\delta Z_{ij}^R/\!\!\!k\omega_+
	-\delta m_{ij}^L\omega_--\delta m_{ij}^R\omega_+
	\nonumber \\
	&&\hspace{1.0cm}=(\delta_{ij}+\delta Z_{ij}^L)\Big(/\!\!\!k-m_i^\text{{tree}}
	-\delta m_{ij}^L+\delta Z_{ij}^Lm_i^\text{{tree}}\Big)\omega_- \nonumber \\
	&&\hspace{1.5cm}+(\delta_{ij}+\delta Z_{ij}^R)\Big(/\!\!\!k-m_i^\text{{tree}}
	-\delta m_{ij}^R+\delta Z_{ij}^Rm_j^\text{{tree}}\Big)\omega_+,
	\label{twopoint}
\end{eqnarray}
where $\delta_{ij}+\delta Z_{ij}^{L}$ is the renormalization
multiplier
for the left-handed wave function and $\delta_{ij}+\delta Z_{ij}^{R}$ is
the renormalization multiplier for the right-handed wave function. $m_i^{tree}$
is the mass of the i-th generation of fermions
at tree level. From Eq.~(\ref{twopoint}) and
the mass-shell conditions, the one-loop correction of the mass matrix elements is obtained as:
\begin{eqnarray}
	&&\delta m_{ij}^\text{{loop}}=\bigg\{\Big[\delta m_{ij}^L
	+\delta m_{ij}^R\Big]_{k^2=0}-\Big[m_i
	\delta Z_{ij}^L|_{k^2=m_i^2}+m_j\delta Z_{ij}^R|_{k^2=m_j^2}
	\Big]\bigg\}\nonumber \\
	&&\hspace{1.cm}=3m_i^\text{{tree}}(m_j^\text{{tree}})^2c_{ij}^1+(m_i^\text{{tree}}m_j^\text{{tree}}
	+(m_i^\text{{tree}})^2+(m_j^\text{{tree}})^2)m_i^\text{{tree}}d_{ij}^1
	\nonumber \\
	&&\hspace{1.5cm}+((m_i^\text{{tree}})^2m_j^\text{{tree}}+3m_i^\text{{tree}}(m_j^\text{{tree}})^2)e_{ij}^1
	+(3(m_i^\text{{tree}})^2m_j^\text{{tree}}+m_i^\text{{tree}}(m_j^\text{{tree}})^2)f_{ij}^1,
	\label{loopmass}
\end{eqnarray}
in which $\delta Z_{ij}^{L,R}$, $\delta m_{ij}^{L,R}$ are defined in
Eq.~(\ref{symbel}). The Eq.~(\ref{loopmass}) is the key formula for calculating the one-loop corrections for the mass matrix of neutrino.

The bosons exchanged in the one-loop self-energy diagrams can be vectors and scalars, and they correspond to different integrals.
The amplitudes for the case of exchanging vector-boson is
\begin{eqnarray}
	&&\text{Amp}_{V}(k)=(\mu_{w})^{2\epsilon}\int\frac{d^DQ}{(2\pi)^D}(iA_{\sigma_1}^{(
		{\tiny V})}\gamma_{\mu}\omega_{\sigma_1})\frac{i(/\!\!\!Q+/\!\!\!k+m_f)}
	{(Q+k)^2-m_f^2}(iB_{\sigma_2}^{({\tiny V})}\gamma^{\mu}\omega_{\sigma_2})
	\frac{-i}{Q^2-m_{{\tiny V}}^2}\nonumber \\
	&&\hspace{1.5cm}=-\int_0^1dx\int\frac{d^DQ}{(2\pi)^D}\frac{1}{(Q^2+x(1-x)k^2
		-xm_f^2-(1-x)m_{\tiny V}^2)^2}\nonumber \\
	&&\hspace{3.0cm}\Big\{(2-D)A_{\sigma}^{({\tiny V})}
	B_{\sigma}^{({\tiny V})}(1-x)/\!\!\!k\omega_{\sigma}
	+Dm_f
	A_{\bar{\sigma}}^{({\tiny V})}B_{\sigma}^{({\tiny V})}\omega_{\sigma}\Big\}
	\nonumber \\
	&&\hspace{1.5cm}=-i\int_0^1dx\int\frac{d^DQ}{(2\pi)^D}\frac{1}{(Q^2
		+xm_f^2+(1-x)m_{\tiny V}^2)^2}\Big\{1+\frac{2x(1-x)k^2}{Q^2
		+xm_f^2+(1-x)m_{\tiny V}^2}\Big\}\nonumber \\
	&&\hspace{2.0cm}\Big\{(2-D)A_{\sigma}^{({\tiny V})}
	B_{\sigma}^{({\tiny V})}(1-x)/\!\!\!k\omega_{\sigma}+Dm_f
	A_{\bar{\sigma}}^{({\tiny V})}B_{\sigma}^{({\tiny V})}\omega_{\sigma}\Big\},
	\label{ampv1}
\end{eqnarray}
where  $D=4-2\epsilon$ and $\mu_w$ represents the
renormalization scale. $A_{\sigma}^{({\tiny V})},\; B_{\sigma}^{
	({\tiny V})}$ with $\sigma=\pm$ are the interaction vertices, which can be got through SARAH.  $m_{\tiny V}$
represents the mass of the vector boson that appears in the loop and
$m_f$ is for the fermion in the loop.
By combining the Eq.(\ref{onel1}), Eq.(\ref{series})
and Eq.(\ref{ampv1}), we get
\begin{eqnarray}
	&&c_{ij}^0(m_{\tiny V},m_f)=-iD\frac{m_f}{m_j}A_{+}^{({\tiny V})}
	B_{-}^{({\tiny V})}F_{2a}(m_f,m_{\tiny V}),\nonumber \\
	&&d_{ij}^0(m_{\tiny V},m_f)=-iD\frac{m_f}{m_i}A_{-}^{({\tiny V})}
	B_{+}^{({\tiny V})}F_{2a}(m_f,m_{\tiny V}),\nonumber \\
	&&e_{ij}^0(m_{\tiny V},m_f)=-i(2-D)A_{-}^{({\tiny V})}
	B_{-}^{({\tiny V})}F_{2b}(m_f,m_{\tiny V}),\nonumber \\
	&&f_{ij}^0(m_{\tiny V},m_f)=-i(2-D)A_{+}^{({\tiny V})}
	B_{+}^{({\tiny V})}F_{2b}(m_f,m_{\tiny V}),\nonumber \\
	&&c_{ij}^1(m_{\tiny V},m_f)=-i4\frac{m_f}{m_j}A_{+}^{({\tiny V})}
	B_{-}^{({\tiny V})}F_{3a}(m_f,m_{\tiny V}),\nonumber \\
	&&d_{ij}^1(m_{\tiny V},m_f)=-i4\frac{m_f}{m_i}A_{-}^{({\tiny V})}
	B_{+}^{({\tiny V})}F_{3a}(m_f,m_{\tiny V}),\nonumber \\
	&&e_{ij}^1(m_{\tiny V},m_f)=i2A_{-}^{({\tiny V})}
	B_{-}^{({\tiny V})}F_{3b}(m_f,m_{\tiny V}),\nonumber \\
	&&f_{ij}^1(m_{\tiny V},m_f)=i2A_{+}^{({\tiny V})}
	B_{+}^{({\tiny V})}F_{3b}(m_f,m_{\tiny V}).
	\label{form1}
\end{eqnarray}
$F_{2a},F_{2b},F_{3a}$ and $F_{3b}$ is the integrals over the
	internal momentum of the loop, and their explicit form is 
\begin{eqnarray}
	&&F_{2a}(m_1,m_2)=(\mu_w)^{2\epsilon}\int_0^1dx\int\frac{d^DQ}{(2\pi)^D}
	\frac{1}{(Q^2+xm_1^2+(1-x)m_2^2)^2}\nonumber \\
	&&\hspace{2.0cm}=\frac{1}{(4\pi)^2}\Big\{\frac{1}{\epsilon}-\gamma_E
	+\frac{m_1^2\ln\frac{4\pi\mu_{w}^2}{m_1^2}-m_2^2\ln\frac{4\pi\mu_{w}^2}
		{m_2^2}}{m_1^2-m_2^2}\Big\},\nonumber \\
	&&F_{2b}(m_1,m_2)=(\mu_w)^{2\epsilon}\int_0^1dx\int\frac{d^DQ}{(2\pi)^D}
	\frac{1-x}{(Q^2+xm_1^2+(1-x)m_2^2)^2}\nonumber \\
	&&\hspace{2.0cm}=\frac{1}{2(4\pi)^2}\Big\{\frac{1}{\epsilon}-\gamma_E
	+\frac{m_1^2-3m_2^2}{2(m_1^2-m_2^2)}\nonumber \\
	&&\hspace{2.5cm}+\frac{1}{(m_1^2-m_2^2)^2}\Big[
	m_1^2(m_1^2-2m_2^2)\ln\frac{4\pi\mu_{w}^2}{m_1^2}+m_2^2\ln\frac{4\pi
		\mu_{w}^2}{m_2^2}\Big]\Big\},\nonumber \\
	&&F_{3a}(m_1,m_2)=(\mu_w)^{2\epsilon}\int_0^1dx\int\frac{d^DQ}{(2\pi)^D}
	\frac{2x(1-x)}{(Q^2+xm_1^2+(1-x)m_2^2)^3}\nonumber \\
	&&\hspace{2.0cm}=\frac{1}{(4\pi)^2}\frac{1}{(m_1^2-m_2^2)^3}\Big\{
	-m_1^2m_2^2\ln\frac{m_1^2}{m_2^2}+\frac{m_1^4-m_2^4}{2}\Big\},
	\nonumber \\
	&&F_{3b}(m_1,m_2)=(\mu_w)^{2\epsilon}\int_0^1dx\int\frac{d^DQ}{(2\pi)^D}
	\frac{2x(1-x)^2}{(Q^2+xm_1^2+(1-x)m_2^2)^3}\nonumber \\
	&&\hspace{2.0cm}=\frac{1}{(4\pi)^2}\frac{1}{(m_1^2-m_2^2)^4}\Big\{
	\frac{1}{3}m_1^6+\frac{1}{6}m_2^6+\frac{1}{2}m_1^4m_2^2-m_1^2m_2^4
	-m_1^4m_2^2\ln\frac{m_1^2}{m_2^2}\Big\}.
	\label{function}
\end{eqnarray}
For the case of exchanging scalar bosons, the amplitudes are derived in a similar way
\begin{eqnarray}
	&&\text{Amp}_{S}(k)=(\mu_{w})^{2\epsilon}\int\frac{d^DQ}{(2\pi)^D}(iA_{\sigma_1}^{(
		{\tiny S})}\omega_{\sigma_1})\frac{i(/\!\!\!Q+/\!\!\!k+m_f)}
	{(Q+k)^2-m_f^2}(iB_{\sigma_2}^{({\tiny S})}\omega_{\sigma_2})
	\frac{i}{Q^2-m_{{\tiny S}}^2}\nonumber \\
	&&\hspace{1.5cm}=i\int_0^1dx\int\frac{d^DQ}{(2\pi)^D}\frac{1}{(Q^2
		+xm_f^2+(1-x)m_{\tiny S}^2)^2}\Big\{1+\frac{2x(1-x)k^2}{Q^2
		+xm_f^2+(1-x)m_{\tiny S}^2}\Big\}\nonumber \\
	&&\hspace{2.0cm}\Big\{A_{\bar{\sigma}}^{({\tiny S})}
	B_{\sigma}^{({\tiny S})}(1-x)/\!\!\!k\omega_{\sigma}+m_f
	A_{\sigma}^{({\tiny S})}B_{\sigma}^{({\tiny
			S})}\omega_{\sigma}\Big\},
	\label{amps1}
\end{eqnarray}
where $A_{\sigma}^{({\tiny S})},\; B_{\sigma}^{({\tiny S})}$ with
$\sigma=\pm$ are the interaction vertices, which can also be obtained through SARAH.  $m_{\tiny S}$
represents the mass of the scalar boson that appears in the loop and
$m_f$ is for the fermion in the loop. From Eq.(\ref{onel1}), Eq.(\ref{series}) and Eq.(\ref{amps1}), we
obtain
\begin{eqnarray}
	&&c_{ij}^0(m_{\tiny S},m_f)=i\frac{m_f}{m_j}A_{-}^{({\tiny S})}
	B_{-}^{({\tiny S})}F_{2a}(m_f,m_{\tiny S}),\nonumber \\
	&&d_{ij}^0(m_{\tiny S},m_f)=i\frac{m_f}{m_i}A_{+}^{({\tiny S})}
	B_{+}^{({\tiny S})}F_{2a}(m_f,m_{\tiny S}),\nonumber \\
	&&e_{ij}^0(m_{\tiny S},m_f)=iA_{+}^{({\tiny S})}
	B_{-}^{({\tiny S})}F_{2b}(m_f,m_{\tiny S}),\nonumber \\
	&&f_{ij}^0(m_{\tiny V},m_f)=iA_{-}^{({\tiny S})}
	B_{+}^{({\tiny S})}F_{2b}(m_f,m_{\tiny S}),\nonumber \\
	&&c_{ij}^1(m_{\tiny S},m_f)=i\frac{m_f}{m_j}A_{-}^{({\tiny S})}
	B_{-}^{({\tiny S})}F_{3a}(m_f,m_{\tiny S}),\nonumber \\
	&&d_{ij}^1(m_{\tiny S},m_f)=i\frac{m_f}{m_i}A_{+}^{({\tiny S})}
	B_{+}^{({\tiny S})}F_{3a}(m_f,m_{\tiny S}),\nonumber \\
	&&e_{ij}^1(m_{\tiny S},m_f)=iA_{+}^{({\tiny S})}
	B_{-}^{({\tiny S})}F_{3b}(m_f,m_{\tiny S}),\nonumber \\
	&&f_{ij}^1(m_{\tiny V},m_f)=iA_{-}^{({\tiny S})}
	B_{+}^{({\tiny S})}F_{3b}(m_f,m_{\tiny S}).
	\label{form2}
\end{eqnarray}

In this work, the mixing of ${\nu_ i}^0-{\nu_ j}^0$ originates from the following loop diagrams.

\begin{itemize}
	\item The internal particles are $Z\sim$gauge boson and neutrinos
	$\nu_{\alpha}^0$ ($\alpha=1,2,\cdots,6$).
	\item  The internal particles are $W\sim$gauge boson and 
	charged leptons $e_\alpha$ ($\alpha=1,2,3$).
	\item The internal particles are CP-even Higgs bosons $H_\beta^0$ ($\beta=1,2,
	\cdots,5$) and  neutrinos $\nu_{\alpha}^0$
	($\alpha=1,2,\cdots,6$).
	\item The internal particles are CP-odd Higgs bosons $A_\beta^0$ ($\beta=1,2,
	\cdots,5$) and neutrinos $\nu_{\alpha}^0$
	($\alpha=1,2,\cdots,6$).
	\item The internal particles are CP-even sneutrinos $\tilde{\nu}^R_\beta$  ($\beta=1,2,
	\cdots,6$) and  neutralinos $\tilde{\chi}_{\alpha}$
	($\alpha=1,2,\cdots,7$).
	\item The internal particles are CP-odd sneutrinos $\tilde{\nu}^I_\beta$ ($\beta=1,2,
	\cdots,6$) and neutralinos $\tilde{\chi}_{\alpha}$
	($\alpha=1,2,\cdots,7$).
	\item The internal particles are charged Higgs bosons $H_\beta^+$ ($\beta=1,2,
	\cdots,4$) and charged leptons $e_\alpha$
	($\alpha=1,2,3$).
	\item The internal particles are slepton $\tilde{e}_\beta$ ($\beta=1,2,
	\cdots,6$) and charginos $\chi_\alpha^-$ 
	($\alpha=1,2,3$).
\end{itemize}
Then, the one-loop corrections to the neutrino mass matrix elements can be obtained
\begin{eqnarray}
	&& \delta m_{\nu{ij}}^{1-\text{loop}}=\delta m_{\nu{ij}}^{(Z,{\nu^0})}+\delta m_{\nu{ij}}^{(W,e^-)}
	+\delta m_{\nu{ij}}^{(H^0,\nu^0)}+\delta m_{\nu{ij}}^{(A^0,\nu^0)} \nonumber\\
	&&\quad \quad+\delta m_{\nu{ij}}^{(\tilde{\nu}^I,\tilde{\chi})}+\delta m_{\nu{ij}}^{(\tilde{\nu}^R,\tilde{\chi})}+\delta m_{\nu{ij}}^{(H^+,e^-)}+\delta m_{\nu{ij}}^{(\tilde{e},\chi^-)}.
	\label{suml}
\end{eqnarray}
\section{Diagonalized the effective neutrino mass matrix\label{TDM}}
Using the "top-down" method \cite{top-down,neu-b5,Yan1} in the effective mass matrix $M_{\nu}^{eff}$, we get the Hermitian matrix
\begin{eqnarray}
	{\cal H}=(M_{\nu}^{eff})^{\dagger}M_{\nu}^{eff}.
\end{eqnarray}

The eigenvalues of the $3\times3$ effective mass squared matrix ${\cal H}$ are given as
\begin{eqnarray}
	&&m_1^2={a\over3}-{1\over3}p(\cos\phi+{\sqrt3}\sin\phi),\nonumber\\
	&&m_2^2={a\over3}-{1\over3}p(\cos\phi-{\sqrt3}\sin\phi),\nonumber\\
	&&m_3^2={a\over3}+{2\over3}p\cos\phi.
	\label{eigenvalues}
\end{eqnarray}
Here the concrete forms of the parameter in Eq.~(\ref{eigenvalues}) is given
\begin{eqnarray}
	&&p=\sqrt{a^2-3b}, ~~~~~\phi={1\over3}\arccos({1\over p^3}(a^3-{9\over2}ab+{27\over2}c)),
	~~~~a={\rm Tr}({\cal H}),\nonumber\\
	&&b={\cal H}_{11}{\cal H}_{22}+{\cal H}_{11}{\cal H}_{33}+{\cal H}_{22}{\cal H}_{33}
	-{\cal H}_{12}^2 -{\cal H}_{13}^2-{\cal H}_{23}^2,~~~~c={\rm Det}({\cal H}).
\end{eqnarray}
For the three neutrino mixing, there are two possible solutions on the neutrino mass spectrum. The normal hierarchy (NH) spectrum is
\begin{eqnarray}
	&&m_{\nu_1}<m_{\nu_2}<m_{\nu_3},\quad m_{\nu_1}^2=m_1^2,\quad m_{\nu_2}^2=m_2^2,\quad m_{\nu_3}^2=m_3^2,\nonumber\\
	&&\Delta{m_{\nu_{21}}^2}=m_{\nu_2}^2-m_{\nu_1}^2,\quad \Delta{m_{\nu_{32}}^2}=m_{\nu_3}^2-m_{\nu_2}^2,
	\label{NO}
\end{eqnarray}
and the inverted hierarchy (IH) spectrum is
\begin{eqnarray}
	&&m_{\nu_3}<m_{\nu_1}<m_{\nu_2},\quad m_{\nu_3}^2=m_1^2,\quad m_{\nu_1}^2=m_2^2,\quad m_{\nu_2}^2=m_3^2,\nonumber\\
	&&\Delta{m_{\nu_{21}}^2}=m_{\nu_2}^2-m_{\nu_1}^2,\quad \Delta{m_{\nu_{32}}^2}=m_{\nu_2}^2-m_{\nu_3}^2.
	\label{IO}
\end{eqnarray}
The orthogonal matrix $U_\nu$ of $\cal{H}$ can be obtained can be obtained from the mass squared matrix $\cal{H}$ and the three eigenvalues \cite{top-down,neu-b5,Yan1}. The mixing angles between three tiny neutrinos can be defined as follows
\begin{eqnarray}
	&&\sin\theta_{13}=\Big|\Big(U_\nu\Big)_{13}\Big|,~~~~~~~~~~~~~~~\cos\theta_{13}=\sqrt{1-\Big|\Big(U_\nu\Big)_{13}\Big|^2},\nonumber\\
	&&
	\sin\theta_{23}={\Big|\Big(U_\nu\Big)_{23}\Big|\over\sqrt{1-\Big|\Big(U_\nu\Big)_{13}\Big|^2}},~~~~~~
	\cos\theta_{23}={\Big|\Big(U_\nu\Big)_{33}\Big|\over\sqrt{1-\Big|\Big(U_\nu\Big)_{13}\Big|^2}},\nonumber\\
	&& \sin\theta_{12}={\Big|\Big(U_\nu\Big)_{12}\Big|\over\sqrt{1-\Big|\Big(U_\nu\Big)_{13}\Big|^2}},~~~~~~
	\cos\theta_{12}={\Big|\Big(U_\nu\Big)_{11}\Big|\over\sqrt{1-\Big|\Big(U_\nu\Big)_{13}\Big|^2}}.
\end{eqnarray}
\section{The Dirac fermions EDM and
	MDM\label{MDM}}
	In fact, all dimensional-6 operators in (\ref{operators}) induce the effective couplings among between fermions and photons, and the vertex containing an external photon can be written as
\begin{eqnarray}
	&&O_1^{L,R} = ie \{((p+k)^2+p^2)\gamma_\rho+({/\!\!\! p}+{/\!\!\! k})\gamma_\rho{/\!\!\! p}\} P_{L,R}, \nonumber\\
	&&O_2^{L,R} = ie ({/\!\!\! p}+{/\!\!\! k})[{/\!\!\! k}, \gamma_\rho] P_{L,R}, \nonumber\\
	&&O_3^{L,R} = ie [{/\!\!\! k}, \gamma_\rho] {/\!\!\! p} P_{L,R}, \nonumber\\
	&&O_4^{L,R} = ie  (k^2\gamma_\rho-{/\!\!\! k}k_\rho) P_{L,R}, \nonumber\\
	&&O_5^{L,R} = ie m_{{\psi}_i} \{({/\!\!\! p}+{/\!\!\! k})\gamma_\rho+\gamma_\rho {/\!\!\! p}\}  P_{L,R}, \nonumber\\
	&&O_6^{L,R} = ie m_{{\psi}_i} [{/\!\!\!k}, \gamma_\rho]P_{L,R}.
\end{eqnarray}
When the fully theory is invariant under the combined transformations of charge conjugation, parity and time reversal (CPT), then the induced effective theory maintains symmetry after the heavy degrees of freedom are integrated out. It implies that the Wilson coefficients of the operator $O_{2,3,6}^{L,R}$ satisfy the relation \cite{Feng}
\begin{eqnarray}
	C_3^{L,R}=C_2^{R,L\ast}, \qquad  C_6^{L}=C_6^{R\ast}\,,
\end{eqnarray}
where $C_I^{L,R}$ ($I=1\cdots6$) represents the Wilson coefficients of the corresponding operator $O_I^{L,R}$ in the effective Lagrangian. Applying the equations of motion to the external fermions, one finds that the relevant terms in the effective Lagrangian change as follows:
\begin{eqnarray}
	&&\quad\; C_2^{R}O_2^R + C_2^{L}O_2^L +  C_2^{L\ast}O_3^R + C_2^{R\ast}O_3^L + C_6^{R}O_6^R + C_6^{R\ast}O_6^L \nonumber\\
	&&\Rightarrow (C_2^{R} + \frac{m_{{\psi}_j}}{m_{{\psi}_i}}C_2^{L\ast} + C_6^{R})O_6^R + (C_2^{R\ast} + \frac{m_{{\psi}_j}}{m_{{\psi}_i}}C_2^{L} + C_6^{R\ast})O_6^L \nonumber\\
	&&=e m_{{\psi}_i} \Re(C_2^{R} + \frac{m_{{\psi}_j}}{m_{{\psi}_i}}C_2^{L\ast} + C_6^{R})\bar{\psi}_i \sigma^{\mu\nu} \psi_j  F_{\mu\nu} \nonumber\\
	&&\quad +\: ie m_{{\psi}_i}\Im(C_2^{R} + \frac{m_{{\psi}_j}}{m_{{\psi}_i}}C_2^{L\ast} + C_6^{R})\bar{\psi}_i \sigma^{\mu\nu} \gamma_5 \psi_j   F_{\mu\nu}.
	\label{CtoRI}
\end{eqnarray}
Where $\Re(\cdots)$ and $\Im(\cdots)$ represent the real and imaginary parts of the chosen complex number, respectively.
Comparing the Eq.~(\ref{MEDM}) and Eq.~(\ref{CtoRI}) , one can obtain
\begin{eqnarray}
	&&\:\epsilon_{ij} = 4m_e m_{{\psi}_i} \Im(C_2^{R} + \frac{m_{{\psi}_j}}{m_{{\psi}_i}}C_2^{L\ast} + C_6^{R}) \mu_{\rm{B}},\nonumber\\
	&&\mu_{ij} = 4m_e m_{{\psi}_i} \Re(C_2^{R} + \frac{m_{{\psi}_j}}{m_{{\psi}_i}}C_2^{L\ast} + C_6^{R}) \mu_{\rm{B}},
	\label{CUU}
\end{eqnarray}
where $\mu_{\rm{B}}=\frac{e}{2 m_e}$ and $m_e$ is the electron mass. Eq.~(\ref{CUU}) indicates that, the EDM and MDM of Dirac fermions are proportional to the imaginary and real parts of the effective coupling $C_2^{R} + \frac{m_{{\psi}_j}}{m_{{\psi}_i}}C_2^{L\ast} + C_6^{R}$, respectively.


\begin{thebibliography}{99}
		
		\bibitem{neu-b1}C. Giunti and C. W. Kim, {\it  Fundamentals of Neutrino Physics and Astrophysics} (Oxford University Press, 2007).
		
		\bibitem{neu-b2}M. C. Gonzalez-Garcia and M. Maltoni, {\it Phys. Rept.} {\bf 460} (2008) 1 [arXiv:0704.1800].
		
		\bibitem{neu-b3}S. Bilenky, {\it  Introduction to the Physics of Massive and Mixed Neutrinos} (Springer, 2010).
		
		\bibitem{neu-b4}Z.-Z. Xing and S. Zhou, {\it  Neutrinos in Particle Physics, Astronomy and Cosmology} (Zhejiang University Press, 2011).
		
    	\bibitem{neu-b5}
		H.~B.~Zhang, T.~F.~Feng, L.~N.~Kou and S.~M.~Zhao,
		Int. J. Mod. Phys. A \textbf{28} (2013) no.24, 1350117.
		\bibitem{up p1} G. F. Giudice and A. Masiero,{\it Phys. Lett. B} {\bf 206, 480}
		(1988).
		\bibitem{hire p1} R. Barbieri and A. Strumia, [arXiv:hep-ph/0007265].
		\bibitem{hire p2} R. Barate et al. (LEP Working Group for Higgs Boson
		Searches, ALEPH Collaboration, DELPHI Collaboration,
		L3 Collaboration, and OPAL Collaboration), {\it Phys. Lett. B}
		{\bf 565, 61} (2003).
		
		\bibitem{NMS p1}U. Ellwanger, C. Hugonie, and A. M. Teixeira, {\it Phys. Rep.}
		{\bf 496, 1 }(2010).
		\bibitem{NMS p2}S. Chang, R. Dermisek, J. F. Gunion, and N. Weiner,
		Annu. {\bf Rev. Nucl. Part. Sci.} {\it58, 75} (2008).
		\bibitem{UH p1}J. D. Mason, Phys. Rev. D 80, 015026 (2009).
		\bibitem{UH p2} U. Ellwanger and C. Hugonie, {\it Mod. Phys. Lett. A} {\bf22,
			1581} (2007).
		\bibitem{UH p3}B. Ananthanarayan and P. Pandita,{\it Phys. Lett. B} {\bf371, 245}
		(1996).
		\bibitem{UH p4} G. Degrassi and P. Slavich,{\it Nucl. Phys. } {\bf B825, 119}
		(2010).
		\bibitem{TM p1} J. Espinosa and M. Quiros,{\it Phys. Lett. B} {\bf279, 92} (1992).
		\bibitem{TM p2} J. Espinosa and M. Quiros, {\it Phys. Lett. B} {\bf 302, 51} (1993).
		\bibitem{TM p3} J. R. Espinosa and M. Quiros,{\it Phys. Rev. Lett.} {\bf 81, 516}
		(1998).
		\bibitem{TNM p1}  K. Agashe, A. Azatov, A. Katz, and D. Kim, {\it Phys. Rev. D}
		{\bf 84, 115024} (2011).
			\bibitem{type1 p1}P. Minkowski, {\it Phys. Lett.} {\bf67B, 421} (1977); M. Gell-Mann,
		P. Ramond, and R. Slansky, Report No. print-80-0576
		(CERN), 1980; T. Yanagida, in Proceedings of the Workshop
		on the Unified Theory and the Baryon Number in the
		Universe, Tsukuba, Japan, 1979, edited by O. Sawada and
		A. Sugamoto (Report No. KEK-79-18, 1979), p. 95; R. N.
		Mohapatra and G. Senjanovic,  {\it Phys. Rev. Lett}. {\bf44, 912}
		(1980); J. Schechter and J. W. F. Valle, {\it Phys. Rev. D} {\bf22, 2227}
		(1980).
		\bibitem{type2 p1}R. N. Mohapatra and G. Senjanovic,  {\it Phys. Rev. D} {\bf23, 165}
		(1981); G. Lazarides, Q. Shafi, and C. Wetterich, {\it Nucl.
			Phys.} {\bf B181, 287} (1981); C. Wetterich, {\it Nucl. Phys}.{\bf B187,
			343} (1981); B. Brahmachari and R. N. Mohapatra, {\it Phys.
			Rev. D} 58, 015001 (1998); R. N. Mohapatra, {\it Nucl. Phys. B},
		Proc. Suppl. {\bf138, 257} (2005); S. Antusch and S. F. King,
		{\it Phys. Lett. B} {\bf597, 199} (2004).
		\bibitem{UPMNS p1} Z. Maki, M. Nakagawa, and S. Sakata, {\it Prog. Theor. Phys.}
		{\bf 28, 870} (1962).
		\bibitem{UPMNS p2}  F. Capozzi, G. L. Fogli, E. Lisi, A. Marrone, D. Montanino,
		and A. Palazzo,{\it Phys. Rev. D } {\bf89, 093018} (2014).
		
		\bibitem{MASS DATA p1}I. Esteban, M. C. Gonz´alez-Garc´ıa, M. Maltoni, T. Schwetz, and A. Zhou, {\it JHEP} {\bf 09} (2020) 178.
 	    \bibitem{XTN1}	M. Agostini et al. (Borexino),  {\it Phys. Rev. D} {\bf96, 091103} (2017).
 	    \bibitem{XTN2}  D. W. Liu et al. (Super-Kamiokande), {\it Phys. Rev. Lett.} {\bf93, 021802} (2004).
 	    \bibitem{XTN3}A. G. Beda, V. B. Brudanin, V. G. Egorov, D. V. Medvedev, V. S. Pogosov, M. V. Shirchenko,
 	    and A. S. Starostin, {\it Adv. High Energy Phys}. 2012, 350150 (2012).
 	    \bibitem{XTN4} H. T. Wong et al. (TEXONO), {\it Phys. Rev. D} {\bf75, 012001} (2007).
 	    \bibitem{XTN5} Z. Daraktchieva et al. (MUNU), {\it Phys. Lett. B} {\bf564, 190} (2003).
 	    \bibitem{XTN6} A. Ayala, I. Dom´ınguez, M. Giannotti, A. Mirizzi, and O. Straniero, {\it Phys. Rev. Lett}. {\bf113,
 	    191302} (2014), 1406.6053.; N. Viaux, M. Catelan, P. B. Stetson, G. Raffelt, J. Redondo, A. A. R. Valcarce, and A. Weiss,
 	    {\it Phys. Rev. Lett}. {\bf111, 231301} (2013), 1311.1669; A. K. Alok, N. R. Singh Chundawat, and A. Mandal, {\it Phys. Lett. B} {\bf839, 137791} (2023),
 	    2207.13034.;A. H. C´orsico, L. G. Althaus, M. M. Miller Bertolami, S. O. Kepler, and E. Garc´ıa-Berro,
 	    JCAP 08, 054 (2014), 1406.6034.
 	    \bibitem{XTN7} E. Aprile et al. (XENON), {\it Phys. Rev. Lett}. {\bf129, 161805} (2022), [arXiv:2207.11330].
 	    \bibitem{XTN8} S.~Singirala, D.~K.~Singha and R.~Mohanta, [arXiv:2307.10898].
 	    \bibitem{xin1}
 	    S.~Singirala, D.~K.~Singha and R.~Mohanta,
 	    Phys. Rev. D \textbf{108}, no.9, 095048 (2023)
 	    doi:10.1103/PhysRevD.108.095048
 	    [arXiv:2306.14801].
 	    \bibitem{onel1}W. Hollik, Fortschr. Phys. {\bf 38}, 165(1990);
 	    A. Denner, Fortschr. Phys. {\bf 41}, 4(1993).
 	    \bibitem{onel2}
 	    T.~F.~Feng and X.~Q.~Li,
 	    Phys. Rev. D \textbf{63}, 073006 (2001)
 	   	    [arXiv:hep-ph/0012300].
 	    \bibitem{yao}J. Liu and Y. P. Yao, Phys. Rev. {\bf D41}, 2147(1990);
 	    H. Simma, D. Wyler, Nucl. Phys. {\bf B344}, 283(1990);
 	    S. Herrlich and J. Kalinowski, Nucl. Phys. {\bf B381}, 50(1992).
 	    \bibitem{Zhao1}
 	    S.~M.~Zhao, T.~F.~Feng, X.~X.~Dong, H.~B.~Zhang, G.~Z.~Ning and T.~Guo,
 	    Nucl. Phys. B \textbf{910}, 225-239 (2016)
 	    doi:10.1016/j.nuclphysb.2016.07.006
 	    [arXiv:1603.09505].
 	    \bibitem{munuSSMOL}P. Ghosh, P. Dey, B. Mukhopadhyaya, et al., J. High Energy Phys. 05 (2010) 087.
	    \bibitem{type12 p1} D. Borah and M. K. Das, \emph{\it Phys. Rev. D} {\bf90, 015006} (2014).
	   	\bibitem{type12-p4}D. Borah and A. Dasgupta, \emph{JHEP}  {\bf11}, 208 (2015),1509.01800.
		\bibitem{type12-p5} D. Borah {\it Nucl. Phys. B} {\bf876} (2013) 575–586. 
		\bibitem{type12-p3} M. K. Das, D. Borah, and R Mishra {\it Phys. Rev. D} {\bf86, 095006} (2012).
		\bibitem{type12-p2}D. Borah and M. K. Das {\it Nuclear Physics} {\bf B 870} (2013) 461–476.
		 \bibitem{top-down}B.~Dziewit, S.~Zajac and M.~Zralek,
		Acta Phys.\ Polon.\ B {\bf 42}, 2509 (2011).
		\bibitem{Yan1}
		Y.~L.~Yan, T.~F.~Feng, J.~L.~Yang, H.~B.~Zhang, S.~M.~Zhao and R.~F.~Zhu,
		Phys. Rev. D \textbf{97}, no.5, 055036 (2018)
		[arXiv:1803.04599].
		\bibitem{CW1} W. Chao, S. Luo, Z.Z. Xing, and S. Zhou, {\it Phys. Rev. D} {\bf77, 016001} (2008).
		\bibitem{CW2} W. Chao, Z.G. Si, Z.Z. Xing, and S. Zhou,  {\it Phys. Lett. B}  {\bf666, 451} (2008).
		\bibitem{CW3} W.~Chao, Z.~g.~Si, Y.~j.~Zheng and S.~Zhou, Phys. Lett. B \textbf{683}, 26-32 (2010).
		\bibitem{EFT1}W. Buchmuler, D. Wyler: {\it Nucl. Phys.} {\bf B268} (1986) p621-653.
		\bibitem{Feng}T.-F.~Feng, L.~Sun and X.-Y.~Yang, {\it Nucl.~Phys.} {\bf B 800} (2008) 221 [arXiv:0805.1122].
		\bibitem{Feng1}T.-F.~Feng, L.~Sun and X.-Y.~Yang, {\it Phys.~Rev. D} {\bf  77} (2008) 116008 [arXiv:0805.0653].
		\bibitem{Feng2}T.-F.~Feng and X.-Y.~Yang, {\it Nucl.~Phys.} {\bf B 814} (2009) 101 [arXiv:0901.1686].
		\bibitem{EFT2}B Grzadkowski , M Iskrzyński ,M Misiak, J Rosiek "Dimension-six terms in the standard model lagrangian." \emph{JHEP}, {\bf 10}(2010)085.
		\bibitem{Broggini}C. Broggini, C. Giunti and A. Studenikin, {\it  Adv. High Energy Phys.} {\bf 2012} (2012) 459526 [arXiv:1207.3980].
		\bibitem{yang}J. L. Yang, H. B. Zhang, C. X. Liu, X. X. Dong and T. F. Feng,
		doi:10.1007/JHEP08(2021)086 [arXiv:2104.03542].
		\bibitem{Zhang1}H.-B. Zhang, T.-F. Feng,  Z.-F. Ge and S.-M. Zhao \emph{JHEP} {\bf 02} (2014) 012.
		\bibitem{NMTM}I. Esteban, M. Gonzalez-Garcia, A. Hernandez-Cabezudo, M. Maltoni and T. Schwetz, JHEP \textbf{01}, 106 (2019).
		\bibitem{PDG}R.L. Workman et al. (Particle Data Group), Prog. Theor. Exp. Phys. 2022, 083C01.
		\bibitem{data1}G. Aad \textit{et al.} [ATLAS], JHEP \textbf{10}, 005 (2020).
		\bibitem{data2}A. M. Sirunyan \textit{et al.} [CMS], JHEP \textbf{04}, 123 (2021).
		\bibitem{data3} G. Aad \textit{et al.} [ATLAS], [arXiv:2106.01676].
	  	\bibitem{nu47}A. de Gouvea and S. Shalgar, JCAP 10 (2012) 027.
		\bibitem{nu48}A. de Gouvea and S. Shalgar, JCAP 04 (2013) 018.
		\bibitem{zy1}
		K.~S.~Babu, S.~Jana and M.~Lindner,
		JHEP \textbf{10}, 040 (2020)
		[arXiv:2007.04291].
		\bibitem{fn1}
		M.~Gozdz, W.~A.~Kaminski, F.~Simkovic and A.~Faessler,
		Phys. Rev. D \textbf{74}, 055007 (2006)
		[arXiv:hep-ph/0606077].
		\bibitem{Yang2}
		J.~L.~Yang, C.~H.~Chang and T.~F.~Feng,
		Commun. Theor. Phys. \textbf{74}, no.8, 085202 (2022).
		\bibitem{zy2} E.~Akhmedov and P.~Mart\'\i{}nez-Mirav\'e,	JHEP \textbf{10}, 144 (2022).
		\bibitem{zy3}
		S.~Jana, Y.~P.~Porto-Silva and M.~Sen,
		JCAP \textbf{09}, 079 (2022).
		
	\end{thebibliography}
\end{document}